\documentclass[fleqn,usenatbib]{mnras}
\usepackage{graphicx}
\usepackage{amssymb}
\usepackage{amsmath}
\usepackage{txfonts}
\usepackage[T1]{fontenc}
\usepackage{ae,aecompl}
\usepackage{cleveref}

\usepackage{pdfsync}
\graphicspath{{figs/}}

\usepackage[dvipsnames]{xcolor}

\title[Dynamical friction and thermal effects]{Numerical study of dynamical friction with thermal effects -- I. Comparison to linear theory}

\author[D. A. Velasco Romero et F. Masset.]{%
David A. Velasco Romero$^{1,2}$\thanks{E-mail: david.velasco@icf.unam.mx} and Fr\'ed\'eric S. Masset$^{2}$\\
$^{1}$Universidad Aut\'onoma del Estado de Morelos, Av. Universidad s/n, 62210 Cuernavaca, Mor., Mexico\\
$^{2}$Instituto de Ciencias F\'isicas, Universidad Nacional Aut\'onoma de M\'exico, Av. Universidad s/n, 62210 Cuernavaca, Mor., Mexico
}

\date{Accepted XXX. Received YYY; in original form ZZZ}

\pubyear{2018}

\begin{document}
\label{firstpage}
\pagerange{\pageref{firstpage}--\pageref{lastpage}}
\maketitle

\begin{abstract}
We investigate by means of numerical simulations with nested Cartesian meshes the force exerted on a massive and luminous perturber moving at constant speed through a homogeneous and opaque gas, taking into account thermal diffusion in the gas and the radiative feedback from the perturber. The force arising from the release of energy into the ambient medium by the perturber, or heating force, is directed along the direction of motion and induces an acceleration of the perturber. Its value is compared to analytic estimates in the low and high Mach number limits, and found to match those accurately. In addition, the drag exerted on a non-luminous perturber significantly departs from the adiabatic expression when thermal diffusion is taken into account. In the limit of a vanishing velocity, this drag tends to a finite value which we determine using linear perturbation theory and corroborate with numerical simulations. The drag on a non-luminous perturber in a non-adiabatic gas therefore behaves like dry or solid friction. We work out the luminosity threshold to get a net acceleration of the perturber and find it to be generally much smaller than the luminosity of accreting low-mass planetary embryos embedded in a gaseous protoplanetary disc at a few astronomical units. We also present in some detail our implementation of nested meshes, which runs in parallel over several \emph{Graphics Processing Units} (GPUs).
\end{abstract}




\defcitealias{2017MNRAS.465.3175M}{MV17}

\section{Introduction}
\label{sec:introduction}
Recent developments in the field of planet-disc interactions show the importance of heat release into the surrounding gas on the dynamical evolution of hot planetary embryos.  The radiative feed back from an embryo yields the appearance of a low density region in its vicinity which can help it withstand or even revert inward migration \citep{2015Natur.520...63B,2017MNRAS.472.4204M}. This low density region can also excite the embryo's eccentricity \citep{2017arXiv170401931E, 2017arXiv170606329C,2017RMxAC..49...15M} and inclination \citep{2017arXiv170401931E,2017RMxAC..49...15M} to sizeable values.

The case of dynamical friction is somewhat simpler, as it does not involve a sheared medium such as a nearly Keplerian disc, but a uniform medium at rest. Yet in this case the release of radiative energy by the perturber into its surroundings also has a strong impact on the back reaction exerted on the perturber, to the point that it may yield a net acceleration, the total force being then a thrust rather than a drag \citep{2017MNRAS.465.3175M,2017ApJ...838..103P}.  \citet[][hereafter MV17]{2017MNRAS.465.3175M} give analytical expressions in the linear regime for the extra force resulting from the radiative feedback (dubbed heating force) in the limit of low and high Mach numbers.  They find that in the low Mach number limit, the heating force has an asymptotic, finite value independent of the perturber's velocity, while in the supersonic regime, the heating force decays as ${\cal M}^{-2}$, ${\cal M}$ being the Mach number. In both cases they find the heating force to be positive.

The aim of the present work is to corroborate these asymptotic behaviours by means of numerical simulations and to investigate the regime of intermediate Mach numbers, for which \citetalias{2017MNRAS.465.3175M} did not provide an analytical estimate. In addition, the expression of the total force requires a knowledge of the net force acting on a non-luminous perturber when thermal diffusion is taken into account. \citetalias{2017MNRAS.465.3175M} made the simple assumption that this force was comparable to that of the adiabatic case, worked out by \citet{1999ApJ...513..252O} and numerically confirmed by \cite{1999ApJ...522L..35S}. We relax here this assumption and study in some detail the force acting on a cold perturber in a medium with thermal diffusion.

A clean numerical ``test bed'' for this problem can be built in the rest frame of the perturber. The gas is initially set to have a velocity opposite to that of the perturber in the gas frame, and it is injected at any instant in time at the upstream boundary with that velocity, while heat diffusion is modelled with a uniform thermal diffusivity over the grid. Owing to the large range of length scales involved in this problem, it is impractical to simulate the system on a single mesh, and we must resort to nested meshes.

\citetalias{2017MNRAS.465.3175M} argue that their linear analysis should be valid when the ratio $GM/\chi c_s$ is small, where $\chi$ is the thermal diffusivity of the gas and $c_s$ the adiabatic sound speed. All the simulations presented here have $GM/\chi c_s\leq 10^{-1}$ (for most of them this ratio is $10^{-2}$). This ratio can be regarded as the ratio between the time scale $R_B^2/\chi$ for diffusion of heat across the perturber's Bondi radius $R_B=GM/c_s^2$ and the acoustic time scale across the Bondi radius\footnote{Throughout this work we consider the Bondi radius  $R_B=GM/c_s^2$ instead of the effective one $R_{B_\text{eff}}=GM/(c_s^2 + V^2)$. Whenever the former is unresolved so is the latter and we therefore only capture the flow in regions where the relative perturbation of density arising from the perturber's gravity is small.}. In the present work, the first of two papers studying the impact of thermal effects and heat release on dynamical friction, we do not attempt to capture the dynamics of the flow within the perturber's Bondi radius, so as to deal only with weakly non-linear flows. Our mesh resolution will therefore be of the order of, or larger than the Bondi radius, and the thermal and acoustic time scales across the Bondi radius are not resolved. Calculations resolving the flow within the Bondi radius and corresponding to arbitrary values of $GM/\chi c_s$ will be presented in the second paper.

In section~\ref{sec:problem-description} we present a description of the problem at hand, remind the main results of \citetalias{2017MNRAS.465.3175M} and introduce the conventions used throughout this paper. We run our simulations with the hydrocode FARGO3D\footnote{\texttt{http://fargo.in2p3.fr}} that we have modified to handle nested meshes. In section~\ref{sec:numerical-implementation} we detail how this new capability of the code was implemented. In section~\ref{sec:results} we present our results, first for the heating force, then for the drag force on a non-luminous perturber in presence of thermal diffusion and finally for the net (total) force on a luminous perturber. In section~\ref{sec:disc-concl} we discuss our findings and give our conclusions.

\section{Problem description}
\label{sec:problem-description}
We introduce a point-like perturber of mass $M$ and luminosity $L$ in an initially uniform gaseous medium with adiabatic sound speed $c_s$. The perturber is kept fixed whilst the gas has initially the uniform velocity $V={\cal M}c_s$. Without loss of generality, we assume the perturber to lie at the origin of a Cartesian frame $(xyz)$ and the gas velocity to be directed along the $z$ axis. The heat released by the perturber in the gas diffuses according to Fourier's law with the uniform and constant thermal diffusivity $\chi$. Resorting to this simple modelling of thermal effects is an oversimple description of the radiative processes at work in the perturber's vicinity. While such a simplification is necessary to make the problem analytically tractable, a numerical treatment could consider a more realistic approach \citep[e.g.][]{2015Natur.520...63B, 2017ApJ...838..103P}. However, sticking to the same simple description of thermal effects as that of the analytical work of \citetalias{2017MNRAS.465.3175M} allows a clean and accurate comparison between analytical expectations and the results of simulations, let aside the moderate computational cost of such prescription.
   
\subsection{Governing equations}
\label{sec:governing-equations}
We denote respectively with $\rho$, $\boldsymbol{v}$ and $e$ the gas density, velocity and density of internal energy, and add a subscript ``$0$'' to refer to unperturbed quantities.
The governing equations are the continuity equation, Euler's equation for the velocity and the energy equation, which read respectively:
\begin{eqnarray}
\label{eq:1}
\partial_t \rho + \nabla \cdot (\rho\boldsymbol{v}) = 0\\
\label{eq:2}
\partial_t \boldsymbol{v} + \boldsymbol{v} (\nabla \cdot \boldsymbol{v}) =  -\frac{\nabla p}{\rho}-\nabla\Phi \\
\label{eq:3}
\partial_t e + \nabla \cdot (e\boldsymbol{v}) + p\nabla \cdot \boldsymbol{v} + \nabla \cdot \left( \chi\rho\nabla \frac{e}{\rho} \right) = L \delta (\boldsymbol{r})
\end{eqnarray} 
where $\Phi= -GM/|\boldsymbol{r}|$ is the perturber's potential and $L\delta(\boldsymbol{r})$ the energy source arising from the radiative feed back on the gas.
In these equations $p=(\gamma-1)e$ represents the pressure and $\gamma$ the adiabatic index.
The fourth term on the left hand side of Eq.~\eqref{eq:3}
arises from thermal diffusion.

\subsection{Conventions used in this work}
\label{sec:conv-used-this}
To avoid ambiguity when necessary, we use the indices $M,L,\chi$ for the solution of Eqs.~\eqref{eq:1} to~\eqref{eq:3}. We also use these indices to refer to the net acceleration $a$ imparted to the perturber:
\begin{equation}
  \label{eq:4}
  a_{M,L,\chi}=\iiint \frac{G\rho_{M,L,\chi}(x,y,z)z}{(x^2+y^2+z^2)^{3/2}}dx\,dy\,dz
\end{equation}
The acceleration $a_{M,0,0}$ is for instance the acceleration imparted to a non-luminous\footnote{Throughout this paper we will indifferently call a perturber with $L=0$ either a non-luminous or a cold perturber.} perturber of mass $M$ in an adiabatic gas ($\chi=0$). When multiplied by $M$, this corresponds to the drag force worked out by \citet{1999ApJ...513..252O}. Similarly, $a_{0,L,\chi}$ is the acceleration imparted to a massless heat source of luminosity $L$ by the perturbation of density triggered in its surroundings. When the relative perturbations of the flow variables are small, they scale linearly with $M$ and $L$, and so does $a_{M,L,\chi}$ by virtue of Eq.~\eqref{eq:4}.

 \begin{figure}
   \includegraphics[width=\columnwidth]{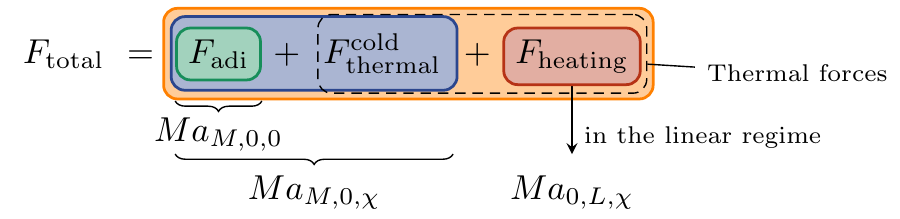}
   \caption{\label{fig:1}Decomposition of the total force acting on the perturber into its adiabatic component and thermal forces (i.e. the forces that can arise when there can be heat exchange between neighbouring fluid elements). Those are further decomposed into the cold thermal force and the heating force. In the electronic version of this manuscript, the colour code also shows the convention that is systematically used in all the graphs representing the force or acceleration in this work: the adiabatic drag is in green, the total force acting on a cold perturber when thermal diffusion is present is represented in blue, the total force in the general case is represented in orange and the heating force in red. For consistency, our naming convention of the forces resembles the naming convention of the torques on a planetary embryo by \citet{2017MNRAS.472.4204M}.}
 \end{figure}

Fig.~\ref{fig:1} summarises the different naming conventions that we will be using throughout this work. The total force acting on a perturber $Ma_{M,L,\chi}$ is decomposed into different components: the adiabatic drag $Ma_{M,0,0}$, the cold thermal force that is the difference between the force exerted on the cold perturber of same mass in a disc with thermal diffusion and an adiabatic disc, and the heating force, which is defined by:
\begin{equation}
  \label{eq:5}
  F_\mathrm{heating} = Ma_{M,L,\chi}-Ma_{M,0,\chi}.
\end{equation}
Only when the flow's perturbation is nearly linear can this force be expressed as $Ma_{0,L,\chi}$.

\subsection{Summary of analytical results}
\label{sec:summ-analyt-results}
We recap hereafter the results obtained by \citetalias{2017MNRAS.465.3175M} that we want to confirm using numerical simulations. \citetalias{2017MNRAS.465.3175M} obtained the expression of the perturbation of density induced by heat release in the limit of a low Mach number, in the linear regime, and derived the corresponding heating force.  They also worked out the expression of the heating force in the case of a large Mach number. All their calculations assume the flow to be in steady state in the perturber's frame.

The perturbation of density due to the heat release given by a linear analysis in the low Mach number limit and in steady state reads, in the case $V>0$:
\begin{equation}\label{eq:6}
 \rho_\mathrm{0,L,\chi}(x,y,z)=-\frac{\gamma(\gamma-1)L}{4\pi\chi c^2_s r}\exp\left(\frac{z-r}{2\lambda}\right),
\end{equation}
where
\begin{equation}
  \label{eq:7}
  \lambda=\chi/\gamma V 
\end{equation}
is the cut-off distance arising from the competition between diffusion and advection and represents the characteristic size of the hot, low density ``plume'' that surrounds the perturber. The acceleration imparted to a test mass in $r=0$ due to this perturbation of density is:
\begin{equation}\label{eq:8} 
a_{0,L,\chi} = \frac{\gamma(\gamma-1)GL}{2\chi c_s^2}\text{sign}(V),  
\end{equation}
This acceleration is directed along the motion of the perturber and is independent on the velocity~$V$. The response time
\begin{equation}
  \label{eq:9}
  \tau = \frac{\lambda^2}{\chi}=\frac{\chi}{\gamma^2V^2}
\end{equation}
is an estimate of the settling time for the heating force.

In the highly supersonic regime, \citetalias{2017MNRAS.465.3175M} give for the acceleration arising from the energy release by the perturber the expression:
\begin{equation}
\label{eq:10}
a_{0,L,\chi} = \frac{2(\gamma-1)GL}{\sqrt{\pi}\chi V^2}f\left(\frac{r_\mathrm{min}V}{4\chi}\right),
\end{equation}
where $r_\mathrm{min}$ is the minimum radius for the force integration and
where $f(\epsilon)$ is given by:
\begin{equation}
  \label{eq:11}
  f(\epsilon)=\int_\epsilon^\infty \mu \left(\sqrt[]{u}\right)du,
\end{equation}
the function $\mu(u)$ being itself defined as:
\begin{equation}
  \label{eq:12}
  \mu(u) = u e^{-u^2}+\frac{\sqrt[]{\pi}}{2}(u^{-2}-2)\text{erfc}(u).
\end{equation}
When the argument of the function $f$ is small (i.e. when the minimal length scale is much smaller than the cut-off distance $\chi/V$), the acceleration can be approximated as:
\begin{equation}
  \label{eq:13}
  a_{0,L,\chi}\approx\frac{(\gamma-1)GL}{\chi V^2}\left[-1.96-\log\left(\frac{r_\mathrm{min}V}{4\chi}\right)\right].
\end{equation}
This acceleration too is along the direction of motion of the perturber. We have designed numerical setups to test Eqs.~\eqref{eq:6}, \eqref{eq:8}, \eqref{eq:10} and~\eqref{eq:13}.

\section{Numerical implementation}
\label{sec:numerical-implementation}
We make use of the \texttt{FARGO3D} code \citep{2016ApJS..223...11B} with a refinement capability through the addition of nested meshes similar to that of the \emph{NIRVANA} code \citep{gda2003}. Our implementation provides the possibility of creating a hierarchy of nested meshes with an arbitrary number of levels of refinement, each finer level doubling the resolution of its coarser predecessor along each dimension. Within this system of nested meshes, each mesh passes individually through the hydrodynamical solver of \texttt{FARGO3D}, the full set of meshes being stitched together by means of inter-level communications at specific stages of the simulation to ensure communicating concurrent information. This nested meshes version of \texttt{FARGO3D} is also implemented in \texttt{CUDA} and \texttt{MPI}, providing the possibility of running in parallel over multiple GPUs. Results already obtained with the \texttt{FARGO3D} code with nested meshes in a different context (that of low mass planets embedded in protoplanetary gaseous discs) have been presented elsewhere \citep{2017AJ....153..124F}. However, the description of our nested mesh implementation in that work was limited. We provide here a more complete description of the different stages required by our implementation. The reader not interested in the implementation details may skip this section and go directly to section~\ref{sec:impl-dynam-frict}.

\subsection{Time step scheduling}
\label{sec:time-scheduling}
In many circumstances the Courant condition implies that the maximum timestep allowed on a given level scales with its resolution. In such a case, for each time step $\Delta t$ on a given level $\ell$, the common practice consists in performing two time steps $\Delta t/2$ on the next level $\ell+1$, in a recursive manner  \citep{gda2002,gda2003,pam08a}. This \emph{sub-cycling} technique is the one we employed in the present work, because it is the most efficient for the problem at hand. However, for different setups such as a planet embedded in a sheared flow, it may not be the best approach. We have developed a general criterion which automatically determines the best sub-cycling scheme for all levels. Since this paper is intended as a reference for our implementation of nested meshes in \texttt{FARGO3D}, we give the detail of our sub-cycling criterion in appendix~\ref{sec:sub-cycl-crit}.

\subsection{Inter-level communications}
\label{sec:inter-level-comm}
In order to coordinate individual meshes into a unique system we need to communicate information in both directions: upwards or coarser to finer, and downwards or finer to coarser. In the upwards direction we have one type of communication (\texttt{CommUp}), in which we communicate data from the coarser level to fill the ghost cells of the finer level. For the downwards direction we have two types of communications: \texttt{CommMean}, where a given number of layers of cells in the coarse level are filled by averaging the underlying finer cells and \texttt{CommFlux}, where the fluxes of the finer level at the boundary between levels are communicated to the coarse level and take precedence
over the fluxes of the coarse mesh.

In the presentation below we consider communications between level $\ell$ (referred to as the coarse or coarser level) and level $\ell+1$ (referred to as the fine or finer level). A full mesh consists of the active cells and the ghost or buffer cells, that surround the active mesh and allow to enforce boundary conditions or communications between meshes. 

\subsubsection{Filling the ghost cells of a finer level: \texttt{CommUp}}
\label{sec:commup}
\begin{figure*}[h]
\includegraphics[width=.95\textwidth]{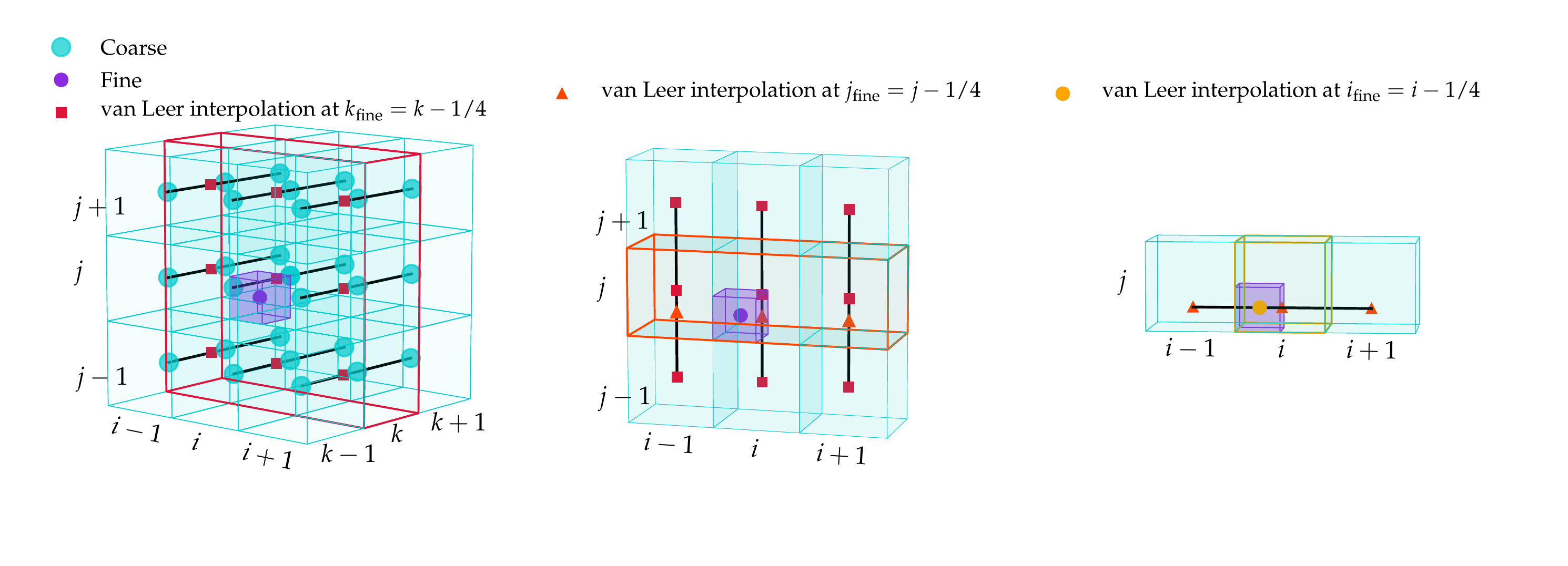}
\caption{Principle of upwards communications (\texttt{CommUp}). We first interpolate over the $z$ direction at the position of the fine cell $k_\mathrm{fine}$, then we proceed with an interpolation over $y$ at $j_\mathrm{fine}$, and we end with an interpolation over $x$ at $i_\mathrm{fine}$. The same process is used for centred and staggered quantities, the only difference between the two being the alignment.}
 \label{fig:2}
\end{figure*}

The ghost cells of a finer level $\ell+1$ are updated before entering the hydrodynamical solver. These ghost cells are filled by a trilinear interpolation of $27$ cells at the coarser level (as shown in fig~\ref{fig:2}). For a fine cell hosted by a coarse cell $(i,j,k)$, the trilinear interpolation starts in the $z$ direction by interpolating values at $k-1$, $k$ and $k+1$ to the position of the fine cell $k_{fine}$, therefore reducing the initial $27$ values into $9$, which are further reduced in the $y$ direction by interpolating values at $j-1$, $j$ and $j+1$ into position $j_{fine}$ to obtain $3$ values at $i-1$, $i$ and $i+1$, which are eventually interpolated at position $i_{fine}$ to obtain the updated value of the fine cell. This sequence is depicted in Fig.~\ref{fig:2}. For each of these interpolations, van Leer's slope limiter is used \citep{1977JCoPh..23..276V}.

This type of communication is needed at the beginning of the cycle on a finer level, which is when both finer and coarser levels are at the same date. Therefore upwards communications are omitted between the two sub-cycles of the finer level in order to avoid communicating outdated information from the coarser level. This requires to double the width of the ghost buffer (its ultimate width depends on the algorithms used).

\subsubsection{Overwriting coarse cells from a fine mesh: \texttt{CommMean}}
\label{sec:commmean}
\begin{figure}
  \begin{center}
    \includegraphics[width=.7\columnwidth]{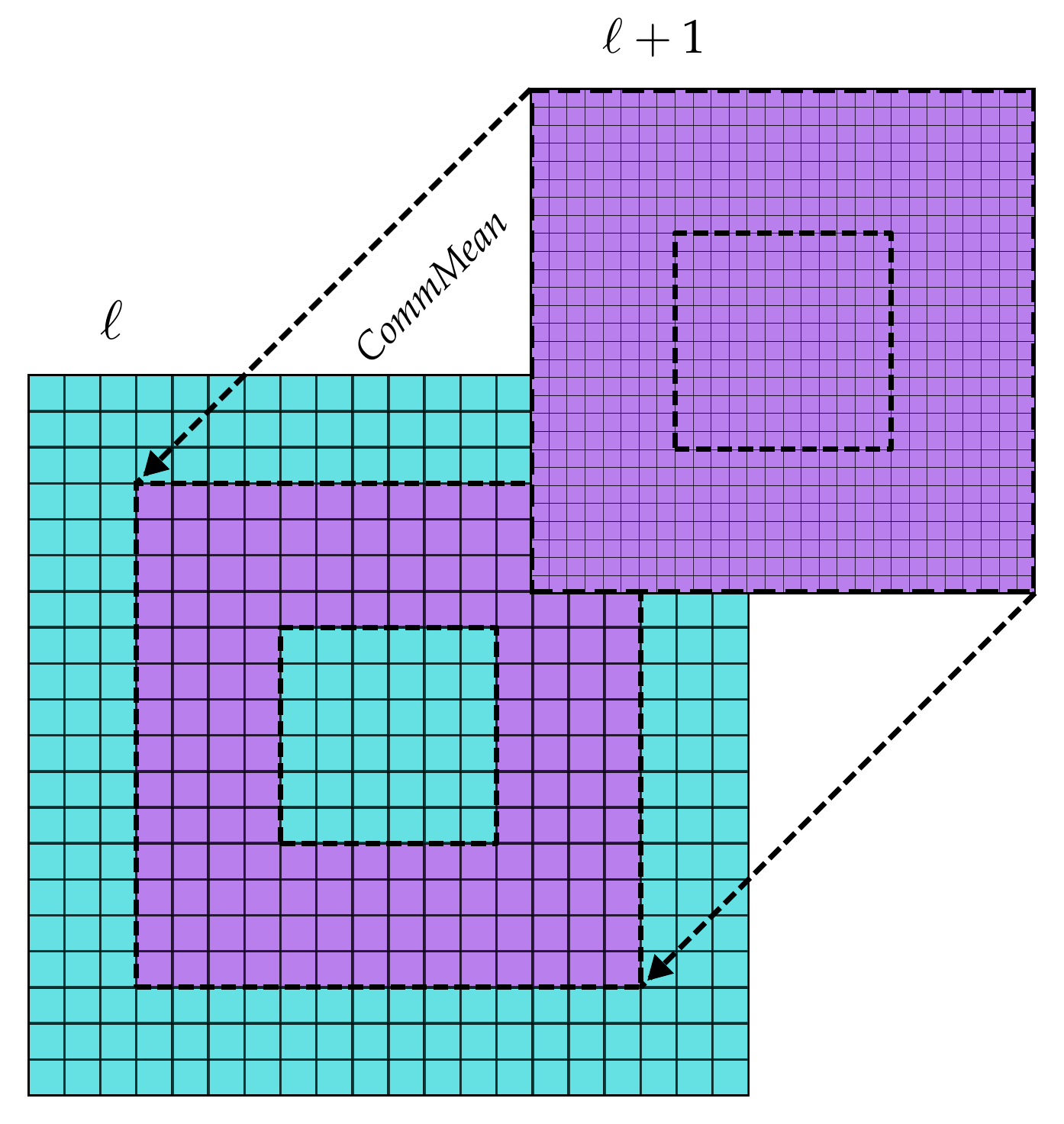}
  \end{center}
  \caption{\label{fig:3}Schematic representation of the  \texttt{CommMean} type of communication. The coarse zones on the inner contour of the fine level $\ell+1$ are filled with the averaged values from that level.}
\end{figure}

\begin{figure}
\includegraphics[width=\columnwidth]{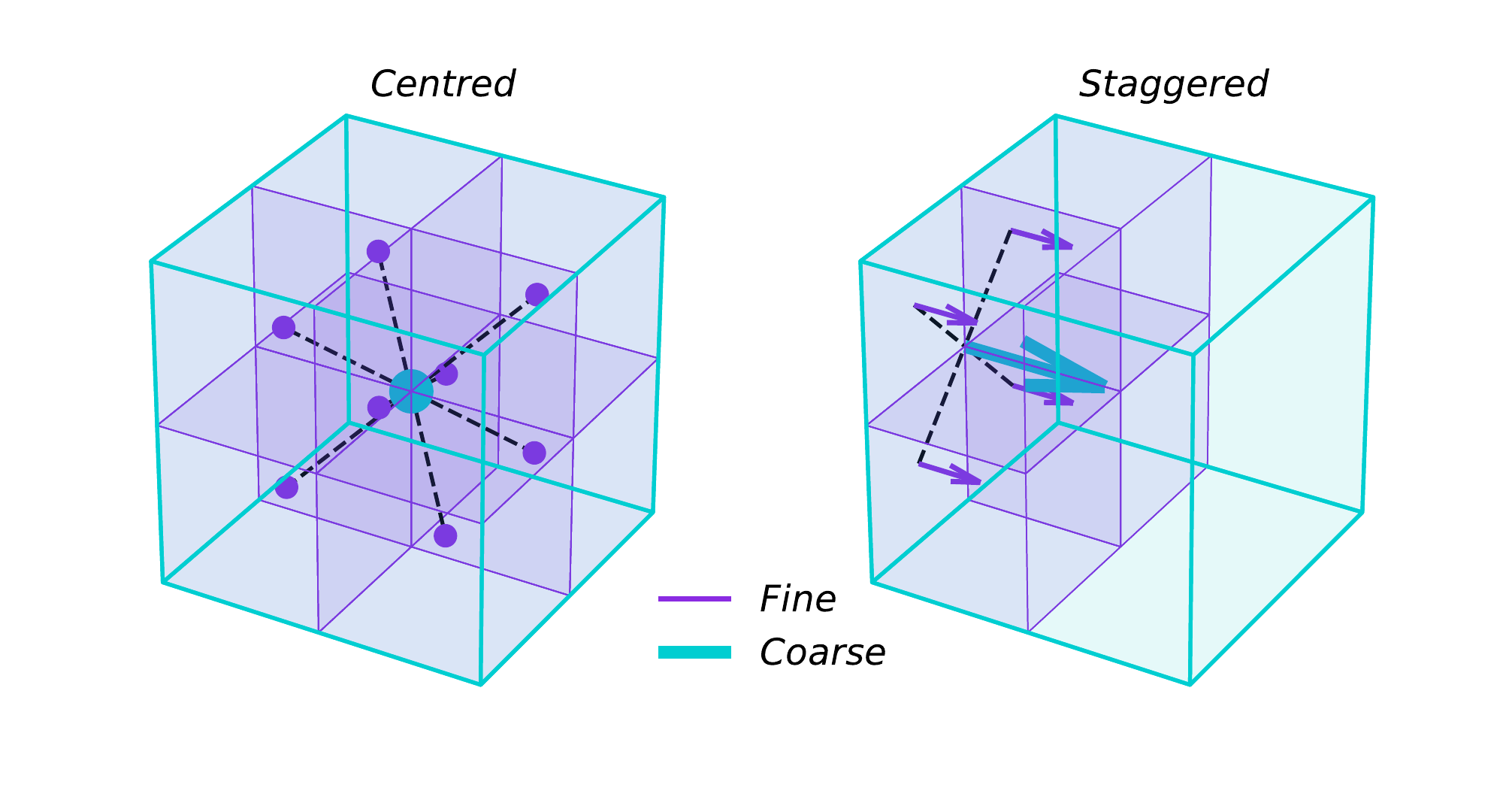}
 \caption{Principle of downwards communications (\texttt{CommDown}). A centred value of a cell in a coarse level is filled by a averaging over $8$ cells of the finer level. Staggered values are obtained by averaging over the $4$ finer cells aligned at the staggered dimension.}
 \label{fig:4}
\end{figure}

At the end of the time step on a coarse level, we overwrite several layers of cells covered by the finer level on the contour of the latter, as depicted in Fig.~\ref{fig:3}. In this implementation it suffices to overwrite four layers given that the number of stages of our solver prevents that information on the coarse level inward of 4 layers reaches the non-refined region. The overwritten cells are filled by averaging the underlying finer cells. Fig~\ref{fig:4} shows that for centred quantities we replace the coarse value with the average of the $8$ finer cells that it contains. For staggered quantities we average over the four finer cells aligned at the staggered dimension.

\subsubsection{Matching fluxes: \texttt{CommFlux}}
\label{sec:commflux}
In order to enforce the conservation of conserved quantities to platform accuracy across all levels we need to match their fluxes at the frontiers between levels. We use the fluxes from the contour of the fine level to determine what goes into the coarse cells that surround this level. In practice this is achieved by sending the fluxes on the contour of level $\ell+1$ to level $\ell$,  and to use these fluxes on the corresponding interfaces of the coarse level
in stead of the fluxes given by the hydrodynamical solver on the coarse level. We communicate and overwrite only fluxes at the frontier.

At a frontier between different levels, the flux $f$ of a coarser cell is therefore overwritten by the sum of the fluxes of the $4$ finer cells that share interface with the coarser cell. The fluxes are communicated at the end of a full cycle of the finer level. In the case of sub-cycling the flux to be communicated is the sum of the fluxes of each sub-cycle $f=f_1+f_2$.

To ensure concurrent fluxes between levels, we have slightly modified the structure of the \emph{source step} described by \citet{2016ApJS..223...11B}, as we have split the so-called source step \emph{SubStep1} (which corresponds to the update of the velocity field under the action of the potential and fictitious forces) into two separate source steps of half the time step at the given level, one before the \emph{transport step} and the other one after it. In stead of the original scheme:
\begin{equation}
  \label{eq:14}
  S(\Delta t)\longrightarrow T(\Delta t), 
\end{equation}
we have the modified scheme:
\begin{equation}
  \label{eq:15}
  S(\Delta t/2) \longrightarrow T(\Delta t)\longrightarrow S(\Delta t/2),
\end{equation}
where $S$ and $T$ stand for the source and transport step, respectively. The evaluation of fluxes is done with the partially updated velocity field $\boldsymbol{v}^p$, obtained upon the end of the first substep.

\begin{figure*}
\includegraphics[width=\textwidth]{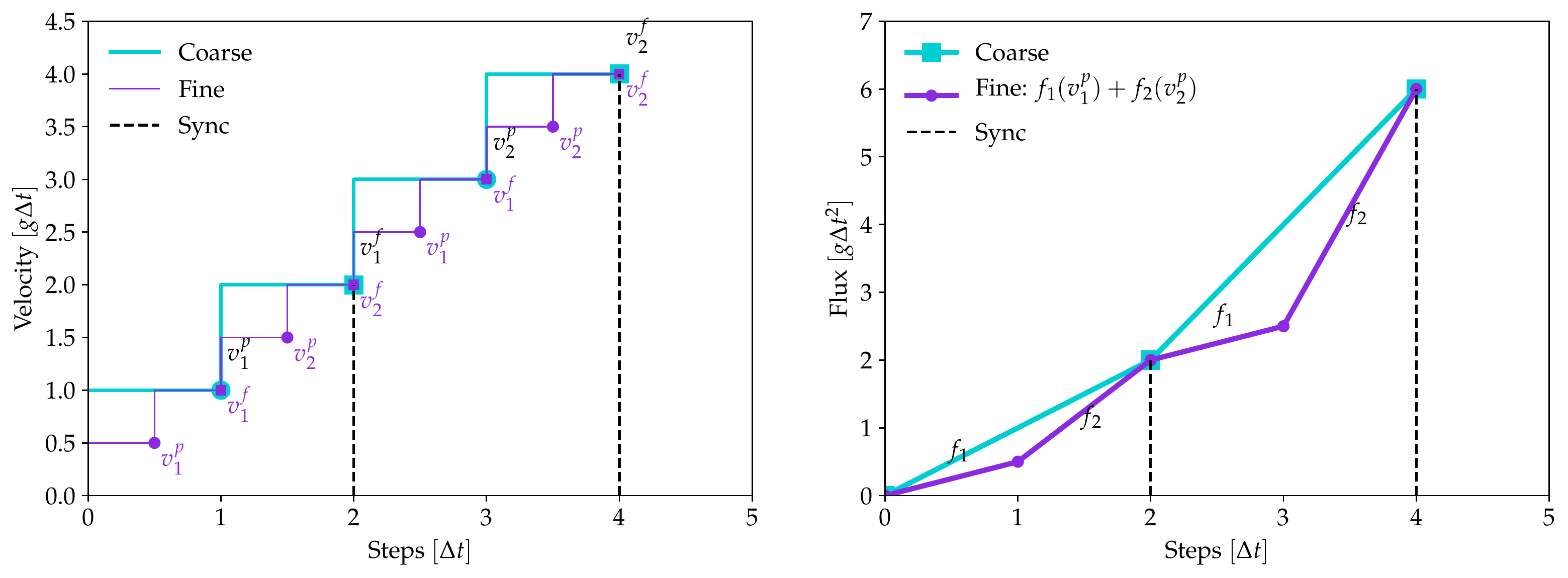}
 \caption{Illustration of the match between velocities and fluxes of coarse and fine levels. The time step of the fine level is $\Delta t$.}
 \label{fig:5}
\end{figure*}

\subsection{Sub-cycling and staging of communications}
\label{sec:sub-cycling-staging}
We present a brief description of our implementation of sub-cycling, showing where the communications are performed. Here we use $\Delta t$ to denote the time step of a sub-cycle on the finer level, we use the superscripts $p$ and $q$ to denote partially updated fields and $f$ for the fully updated fields, while we use subscripts to number the sub-cycles. The names of the different substeps are those of \citet{zeus} and \citet{2016ApJS..223...11B}. 

\begin{itemize}
	\item Finer level $(\ell+1)$:
	\begin{itemize}
    	\item First Sub-cycle:
		\begin{itemize}
        	\item \emph{SubStep1}$(\Delta t/2) \rightarrow \boldsymbol{v}^p_1$.
        	\item \emph{SubStep2,SubStep3}$(\Delta t)\rightarrow \{\rho,e\}^p_1$ 
        	\item \emph{FluxIntegration}$(\boldsymbol{v}^p_1,\Delta t) \rightarrow f_1$
 			\item \emph{Transport}$(f_1) \rightarrow \{\rho,e\}^f_1,\boldsymbol{v}^q_1$.
			\item \emph{SubStep1}($\Delta t/2) \rightarrow \boldsymbol{v}^f_1$.
		\end{itemize}
        \item Second Sub-cycle:
		\begin{itemize}
			\item \emph{SubStep1}($\Delta t/2) \rightarrow \boldsymbol{v}^p_2$.
			\item \emph{SubStep2,SubStep3}$(\Delta t)\rightarrow \{\rho,e\}^p_2$
        	\item \emph{FluxIntegration}$(\boldsymbol{v}^p_2,\Delta t) \rightarrow f_2$
 			\item \emph{Transport}$(f_2) \rightarrow \{\rho,e\}^f_2,\boldsymbol{v}^q_2$.
			\item \emph{SubStep1}($\Delta t/2) \rightarrow \boldsymbol{v}^f_2$.
      \end{itemize}
       	\item \emph{CommFlux}: $f_1+f_2 \rightarrow $ Coarse level
        
	\end{itemize}
    
	\item Coarse level $(\ell)$:
    \begin{itemize}
    	\item First Cycle:
		\begin{itemize}
    		\item \emph{SubStep1}($\Delta t) \rightarrow \boldsymbol{v}^p$.
        	\item \emph{SubStep2,SubStep3}$(2\Delta t)$ 
			\item \emph{FluxIntegration}$(\boldsymbol{v}^p,2\Delta t) \rightarrow f$.
            \item \emph{OverWriteFluxes}$(f_1+f_2) \rightarrow f$.
 			\item \emph{Transport}$(f) \rightarrow \{\rho,e\}^f,\boldsymbol{v}^q $.
        	\item \emph{SubStep1}($\Delta t) \rightarrow \boldsymbol{v}^f$.
        	\item \emph{CommUp}$(\ell \rightarrow \ell+1)$: $\{\rho,\boldsymbol{v},e\}^f \rightarrow \{\rho,\boldsymbol{v},e\}^f_2 $
    		\item \emph{CommDown}$(\ell \leftarrow \ell+1)$:       
            $\{\rho,\boldsymbol{v},e\}^f \leftarrow \{\rho,\boldsymbol{v},e\}^f_2$
            
        \end{itemize}
    \end{itemize}
    
\end{itemize}
We then have concurrent velocities and concurrent fluxes at the end of a coarse cycle, as shown in figure~\ref{fig:5}.

\subsection{Implementation of the dynamical friction setup}
\label{sec:impl-dynam-frict}
We have built a setup with Cartesian meshes where a perturber of mass $M$ and luminosity $L$ is kept fixed at $\boldsymbol{r_0}=(0,0,0)$. The perturber is immersed in a gas with an initially uniform density $\rho_0=1$, adiabatic sound speed $c_s=1$ and an adiabatic index $\gamma=1.4$. The initial energy is then $e_0=c_s^2\rho_0/\gamma(\gamma-1)$. The gas is advected at constant velocity $V$ in the $z$ direction and has a constant thermal diffusivity $\chi$. Taking advantage of the symmetry of the problem in the $x$ and $y$ directions, we reduce our spatial domain to the positive quadrant $x>0$, $y>0$. The ground or coarsest level $\ell=0$ sets the main domain with $x_{min}=0$, $x_{max}=l$, $y_{min}=0$, $y_{max}=l$, $z_{min}=-l$ and $z_{max}=l$, where the extension $l$ is chosen to fulfil a number of conditions that we specify below. Each nested mesh of level $\ell$ covers the sub-domain: $x_{\ell}\in [0, 2^{-\ell} x_{max}]$, $y_{\ell}\in [0, 2^{-\ell} y_{max}]$ and $z_{\ell}\in [2^{-\ell} z_{min}, 2^{-\ell} z_{max}]$.

The gravitational potential follows Plummer's law:
\begin{equation}
  \label{eq:16}
  \Phi(\boldsymbol{r}_{i,j,k})=\frac{GM}{\sqrt[]{|\boldsymbol{r}_{i,j,k}-\boldsymbol{r}_0|^2+\epsilon^2},}
\end{equation}
where $\epsilon$ is the softening length. As said in the introduction, we restrict ourselves in the present work to the study of perturbations with a small relative amplitude, so as to compare our results to those given by the linear analysis of \citetalias{2017MNRAS.465.3175M}. As a consequence, we renounce to describe the dynamics within the perturber's Bondi radius
\begin{equation}
  \label{eq:17}
  R_B=GM/c_s^2, 
\end{equation}
and adopt for the cells of the finest level a size comparable to or larger than $R_B$. A study involving much more resolved calculations showing the flow within the Bondi sphere will be presented in a forthcoming work. Because of this limitation which implies that the perturbation of density is not large even in the vicinity of the perturber, and because the perturber lies on the vertex of the mesh in all our calculations, we generally use $\epsilon=0$, unless explicitly specified.

We monitor the $z$ component of the force that the gas exerts onto the perturber
\begin{equation}
  \label{eq:18}
F= \sum_l \frac{z_lGM\rho_lV_l}{(r'^2_l+\epsilon^2)^{3/2}},
\end{equation}
where $r'_l=|\boldsymbol{r}_l-\boldsymbol{r}_0|$ is the distance from the centre of cell $l$ to the perturber. In practice evaluating this sum over all the cells is done level by level, and care has to be taken to discard the values of coarse zones covered by a finer level.

Thermal diffusion is implemented by evaluating the energy fluxes given by Fourier's law. Upon multiplication by the surface area of the interface they are associated to (either $S_x$, $S_y$ or $S_z$), they read:
\begin{eqnarray}
  \label{eq:19}
  f^x_{i-\frac12} = \chi\rho_0 \left(\frac{e}{\rho}-\frac{e_{i-1}}{\rho_{i-1}}\right)\frac{S_x}{\Delta x}\\
  \label{eq:20}
  f^y_{j-\frac12} = \chi\rho_0 \left(\frac{e}{\rho}-\frac{e_{j-1}}{\rho_{j-1}}\right)\frac{S_y}{\Delta y}\\
  \label{eq:21}
  f^z_{k-\frac12} = \chi\rho_0 \left(\frac{e}{\rho}-\frac{e_{k-1}}{\rho_{k-1}}\right)\frac{S_z}{\Delta z}.
\end{eqnarray}
In the above expressions, we only write indices which differ respectively from $i$, $j$ or $k$, for the sake of legibility.
The energy is then updated as:
\begin{equation}
  \label{eq:22}
	e_{i,j,k} -= \Delta t \left(f^x_{i+\frac12}-f^x_{i-\frac12}+ f^y_{j+\frac12}-f^y_{j-\frac12}+f^z_{k+\frac12}-f^z_{k-\frac12} \right)V^{-1},
\end{equation}
where $V$ represents the volume of the cell.

We implement the release of heat by the luminous point-like perturber in a much similar manner to that of \citet{2015Natur.520...63B}: since the perturber falls exactly at a mesh vertex, the energy $L\Delta t$ released
by the perturber over a time step is split equally between the eight adjacent zones and used to increment their internal energy. 

The boundary conditions of our setup are as follows:
\begin{itemize}
\item In $x_{min}$ and $x_{max}$:
	\begin{itemize}
	\item For $\rho,v_y,v_z,e$, we use symmetric boundary conditions.
    \item For $v_x$, we use antisymmetric boundary conditions.	
	\end{itemize}
\item In $y_{min}$ and $y_{max}$:
	\begin{itemize}
	\item For $\rho,v_x,v_z,e$, we use symmetric boundary conditions.
    \item For $v_y$, we use antisymmetric boundary conditions.	
	\end{itemize}
\item In $z_{min}$ and $z_{max}$:
  \begin{itemize}
    \item
	For $\rho,v_z,e$, we use in/out-flow boundary conditions (we set these variables respectively to $\rho_0,V,e_0$ in the ghost zones).
    \item For $v_x,v_y$, we use symmetric boundary conditions.	
	\end{itemize}	
\end{itemize}

\section{results}
\label{sec:results} All simulations presented here were run up to $t_\mathrm{end}=100\tau$, where $\tau$ is given by Eq.~\eqref{eq:9} and depends on the wind's speed $V$. We chose this duration to ensure that the flow relaxes to a nearly steady state.  For the calculations involving the net force, we take care that the acoustic sphere triggered by the insertion of the perturber in the gas does not reach the edge of the coarser mesh over the duration of the simulation\footnote{When this sphere is reflected on the outer edge of the mesh, the net force exerted on the perturber changes and does no longer represent the dynamical friction.}. Calculations focusing specifically on one of the thermal forces are obtained by subtracting the outcomes of two different calculations\footnote{A calculation with a luminous perturber and a calculation with a non-luminous perturber for the heating force, and a calculation with thermal diffusion and an adiabatic one for the cold thermal force.}. Our calculations being in the nearly linear regime, a reflection of the acoustic sphere on the edge of the computational domain does not have a measurable impact on the thermal forces. For reasons of computational cost we occasionally adopt a smaller mesh, on the edges of which the acoustic sphere is reflected in the course of the simulation, when focusing exclusively on thermal forces.


We quote the mass values in units of $c_s^3/\sqrt{G^3\rho_0}$, those of thermal diffusivity in units of $c_s^2/\sqrt{G\rho_0}$ and those of luminosity in units of $c_s^5/G$.  For the sake of legibility, we will not write explicitly the units after each value that we quote. They correspond to the actual values when $c_s = 1$, $G=1$ and $\rho_0=1$, which is our choice in the simulations.   This scale free unit system allows to capture indistinctly the dynamical friction on a black hole in a gaseous medium or on a proto-planetary embryo in a proto-planetary disc.

\subsection{Heating force}
\label{sec:heating-force}
\begin{figure}
\includegraphics[width=\columnwidth]{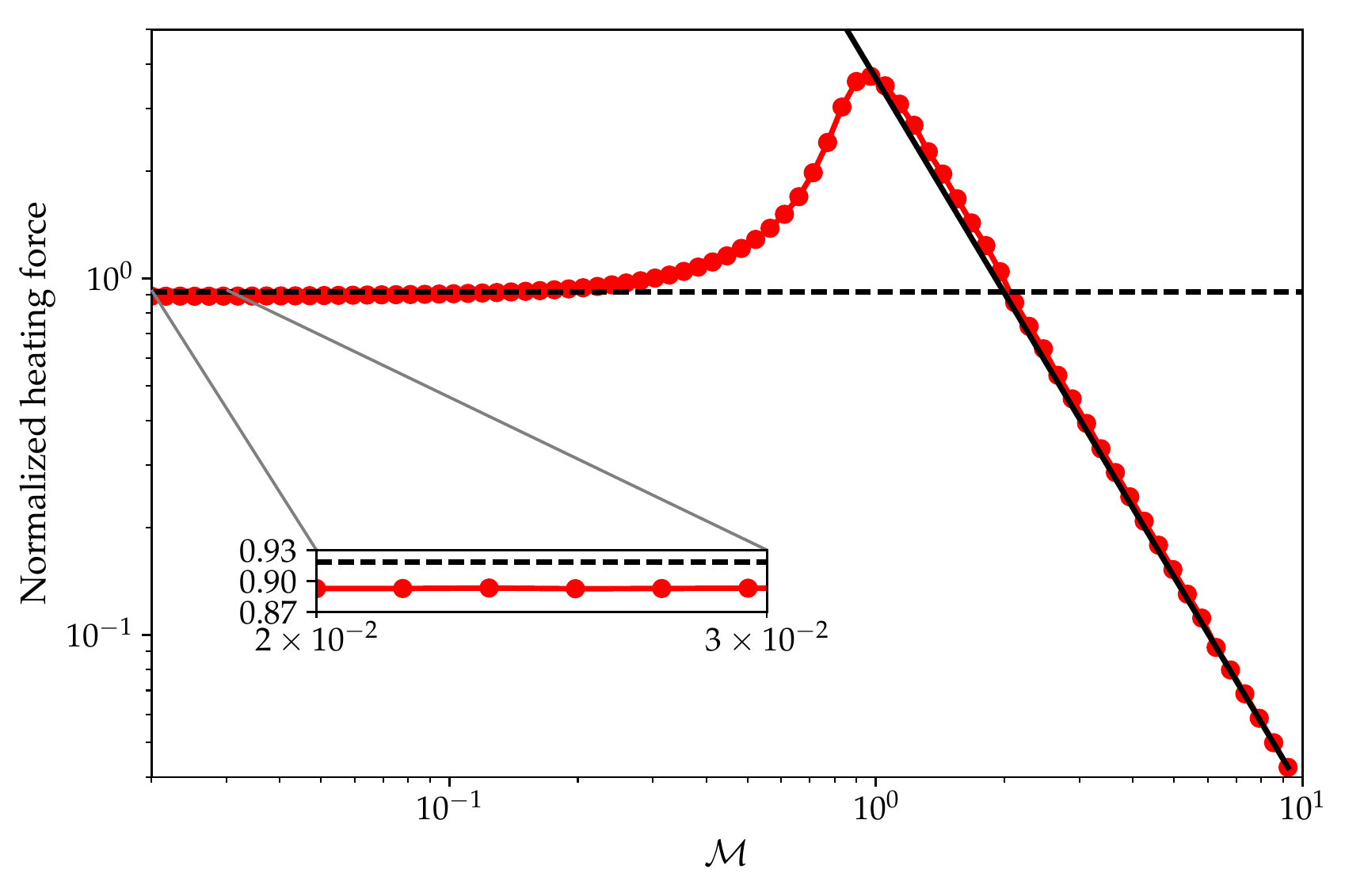}
\caption{Heating force as a function of Mach number, normalized to its theoretical asymptotic value in the ${\cal M}\rightarrow 0$ limit. The horizontal dashed line shows the asymptotic value $h(r'_\mathrm{max})-h(r'_\mathrm{min})$ expected at low Mach number for our finite domain. The inset plot shows that we obtain a force within $\sim 3\%$ of the expected value. The solid line shows an ${\cal M}^{-2}$ law that passes through the rightmost data point.}
 \label{fig:6}
\end{figure}

We undertook a set of simulations spanning 80~values of the Mach number $\cal{M}$ $\in [0.02,5]$ in a geometric sequence, with thermal diffusivity $\chi=0.1$ and a perturber of mass $M=10^{-3}$. For each Mach number, two calculations were run: one with luminosity $L=2\cdot 10^{-4}$ and another one with a cold perturber.  The heating force is obtained by subtracting the forces of the cold runs from the forces of the hot runs.  In this set of calculations we used a box with $l=100\lambda$ thus allowing acoustic waves to bounce off the boundaries for low Mach numbers (in which case $100\tau c_s>100\lambda$). The heating force being the result of a subtraction, it is impervious to this acoustic bounce for flows with nearly linear perturbations such as the ones we consider here. We use here $6$ mesh levels in total, including the ground one, each mesh with sizes $N_x=16$, $N_y=16$ and $N_z=32$. Fig.~\ref{fig:6} shows the heating force as a function of the Mach number. It confirms expectations for the two regimes studied by \citetalias{2017MNRAS.465.3175M}: a finite, asymptotic value at low Mach number, and a force inversely proportional to the speed squared in the supersonic regime. The heating force is positive in all cases. We study in more details these two regimes below.

\subsubsection{Low Mach number}
\label{sec:low-mach-number}
In the limit of a low Mach number, the relative contribution to the net heating force from all the material within a radius $r$ is \citepalias{2017MNRAS.465.3175M}:
\begin{equation}
  \label{eq:23}
  h(r') = \frac{1-2r'-e^{-2r'}}{2r'²}+1,
\end{equation}
where $r'=r/2\lambda$. The fraction of the net force that we can obtain in a simulation is therefore $\approx h(r'_\mathrm{max})-h(r'_\mathrm{min})$, where $r'_\mathrm{max}$ corresponds to the maximal length scale (the box size) and $r'_\mathrm{min}$ corresponds to the minimal length scale (the resolution of the finest level).

For the simulations presented in figure~\ref{fig:6}, the finest resolution amounts to $\Delta x = \frac{25}{128}\lambda$, therefore $r'_{min}=\frac{25}{256}\approx 9.8\cdot 10^{-2}$ and $r'_{max}=l/2\lambda=50$.  We therefore expect to get $h(r'_{max})-h(r'_{min})\approx 92$\% of the full heating force due to our finite length scale range. This is consistent with the force that we obtain in the low Mach number limit, as can be seen in the inset of Fig.~\ref{fig:6}.

\begin{figure}
  \centering
  \includegraphics[width=\columnwidth]{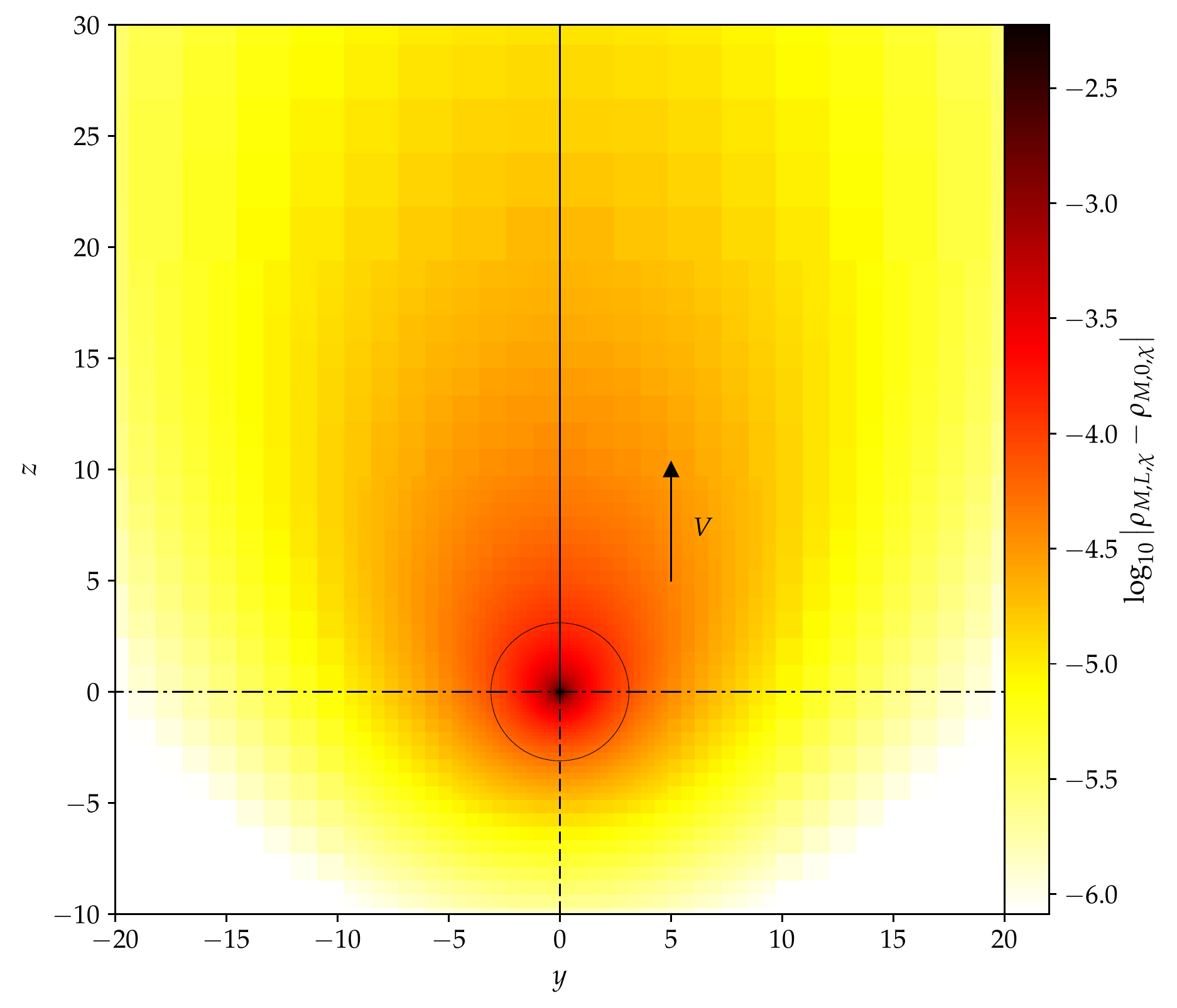}
  \caption{Perturbation of density arising from the release of heat in a plane containing the trajectory of a perturber with a low Mach number. The circle centred on the perturber has radius $\lambda$. Within this circle, the perturbation tends to have a spherical symmetry, whereas at larger radius it is distorted by the flow. The several lines going through the perturber have a style that matches that of the cuts shown in Fig.~\ref{fig:8}. The arrow depicts the direction of the velocity of the gas in the perturber's frame.}
  \label{fig:7}
\end{figure}

\begin{figure}
\includegraphics[width=\columnwidth]{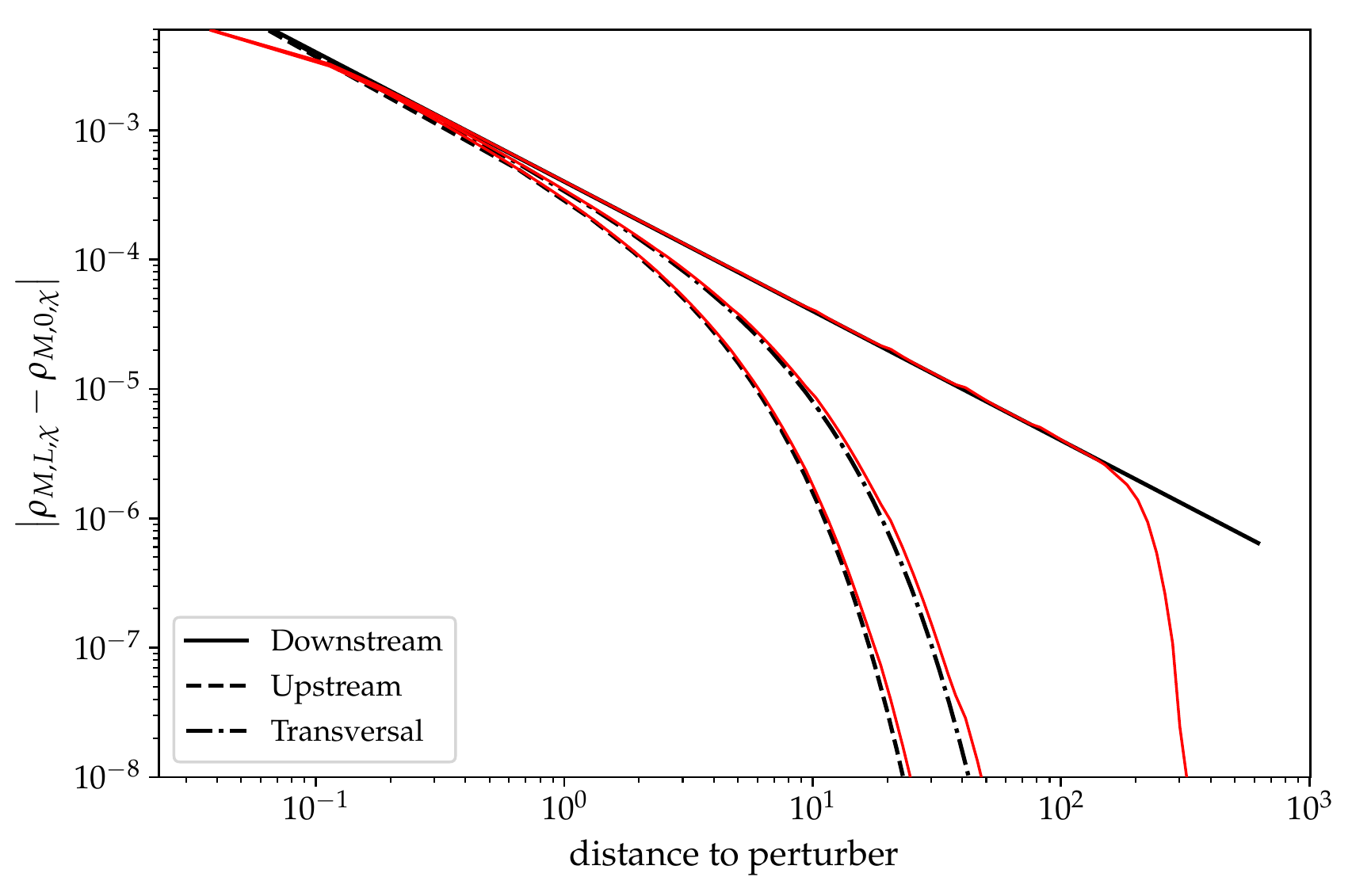}
 \caption{Profile of the density perturbation at low Mach number. The black curves represent the analytic expectations. The perturbation downstream of the perturber does not depend on the wind's speed and simply follows an inverse law \citepalias{2017MNRAS.465.3175M}, whereas it is cut-off in the transverse and upstream directions, the cut-off length being twice shorter in the latter case than in the former.}
 \label{fig:8}
\end{figure}
In the low Mach number limit, the density perturbation is described by Eq.~\eqref{eq:6}. In Fig.~\ref{fig:7} we show a map of the density perturbation arising from the heat release and in Fig.~\ref{fig:8} we plot three different cuts of the density perturbation: along, against and perpendicularly to the direction of advection ($z$) for the case ${\cal M}=0.023$ with a ground level of size $64\times64\times128$ and $11$ levels of refinement of size $32\times32\times64$. Here the first level covers a quarter of the ground level in a given direction, in stead of half of the ground level, which allows to cover a larger domain. We observe a good agreement between numerical results and the analytic expectation. Discrepancies are observed only near the boundaries, where the perturbation of density is negligible, and at distances to the perturber comparable to the resolution of the finest level.

\subsubsection{High Mach number}
\label{sec:high-mach-number}
\begin{figure}
\includegraphics[width=.95\columnwidth]{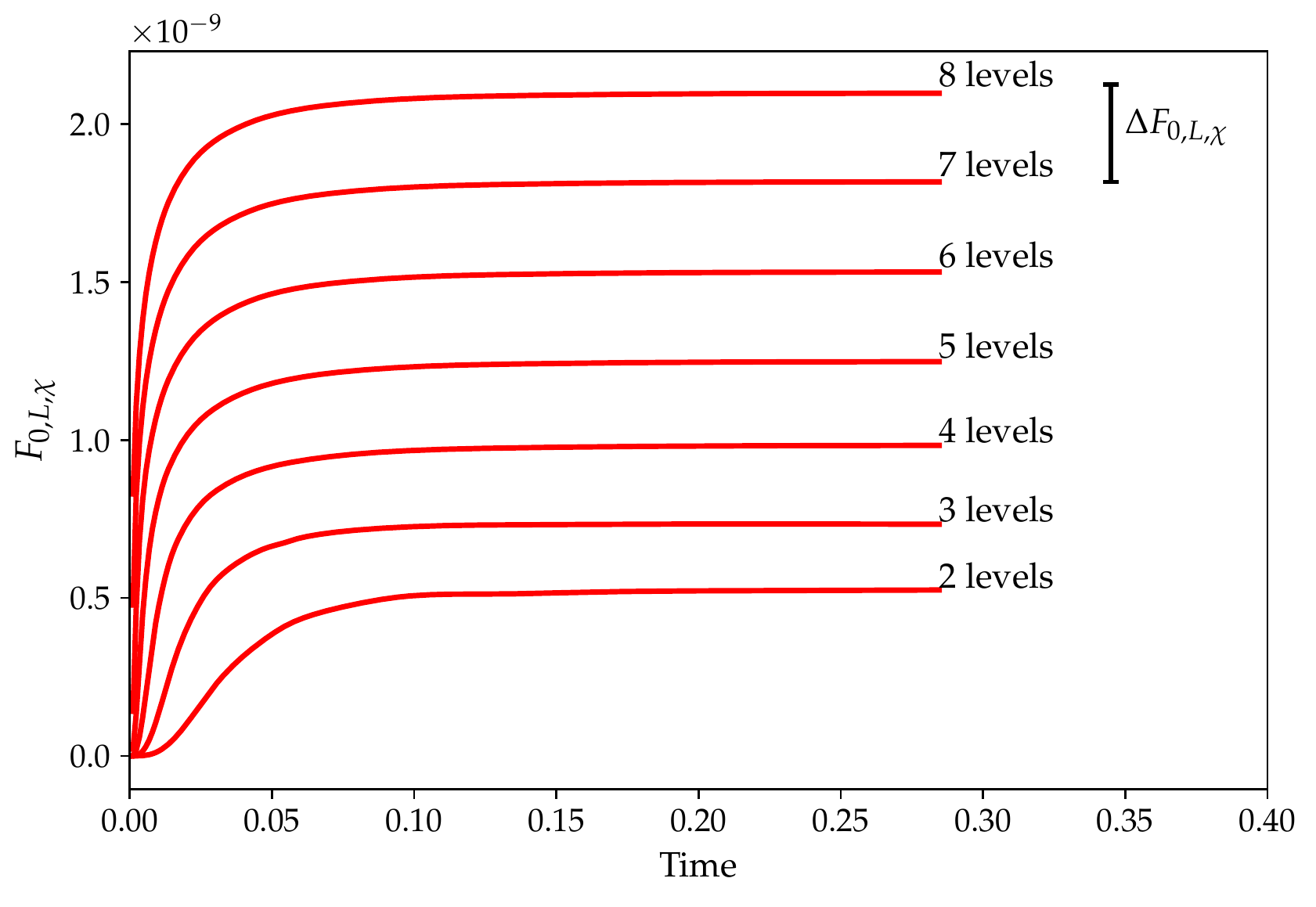}
 \caption{Heating force as a function of time for a perturber with ${\cal M}=3$, for different resolutions. The resolution is doubled for each level added. The vertical bar represents the increment expected for a doubling of resolution given by Eq.~\eqref{eq:24}.}
 \label{fig:9}
\end{figure}

\begin{figure}
  \centering
  \includegraphics[width=\columnwidth]{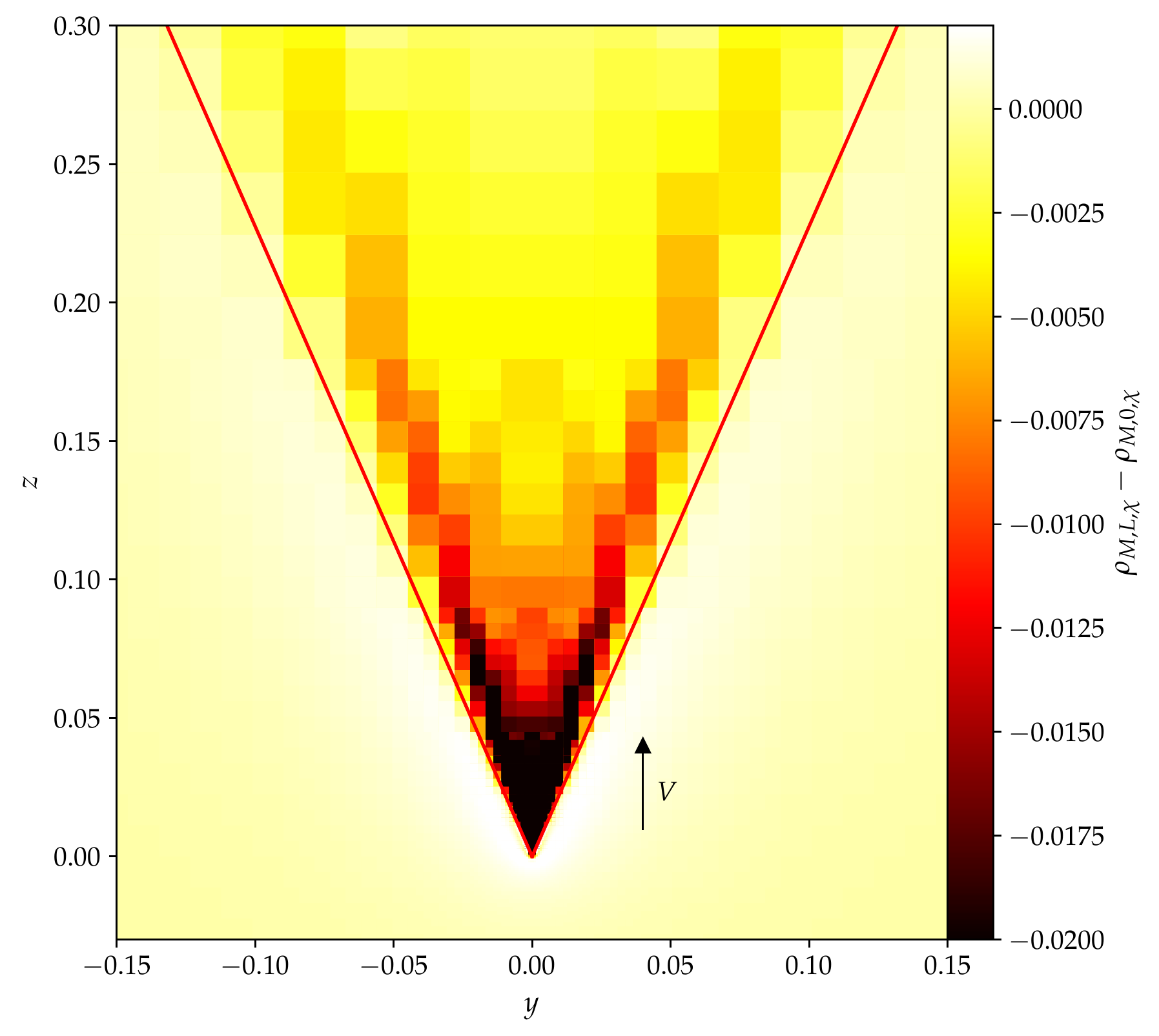}
  \caption{\label{fig:10}Perturbation of density arising from the heat release, in a plane containing the perturber's trajectory, for ${\cal M}=3$. The arrow depicts the gas velocity in the perturber's frame and the red lines delineate the Mach cone.}
\end{figure}

As said above, our setups are such that $l/\lambda$ is fixed, and so is therefore $r_\mathrm{min}V/4\chi\equiv r_\mathrm{min}/(\lambda\times 4\gamma)$. The only dependence on velocity of the heating force in the supersonic regime, given by Eq.~\eqref{eq:10}, is thus through the denominator in $V^{-2}$, which is in agreement with the plot of Fig.~\ref{fig:6}. We further investigate the dependence of that force on the resolution, or $r_\mathrm{min}$, for a fixed Mach number in the supersonic regime. We reproduce a number of times the same calculation, adding each time one level of refinement. The smallest scale is therefore halved from one simulation to the next. In the limit where $r_\mathrm{min}V/4\chi$ is sufficiently small, we see from Eq.~\eqref{eq:13} that the variation of the force between two successive calculations should be: \begin{equation}
  \label{eq:24}
  \Delta F_{0,L,\chi}=\frac{(\gamma-1)GML}{\chi V^2}\log 2.
\end{equation}
This increment is in excellent agreement with the results depicted in Fig.~\ref{fig:9}.
The density perturbation arising from the release of heat exhibits the conical shape typical of the supersonic case, as shown in Fig.~\ref{fig:10}. The perturbation, however, is not strictly bound to the interior of the Mach cone in the vicinity of the apex: thermal diffusion has a typical spread in $\sqrt{\chi t}$ and is therefore ``infinitely fast'' in the vicinity of the perturber, allowing it to beat the acoustic perturbations that define the cone.

\subsubsection{Intermediate Mach number values}
\label{sec:interm-mach-numb}
The analytic results of \citetalias{2017MNRAS.465.3175M} have been established both for the low and high Mach number regimes. No general expression has been given for intermediate values of the Mach number. Fig.~\ref{fig:1} shows that the ${\cal M}^{-2}$ dependence in the supersonic regime is valid almost until ${\cal M}=1$. In Fig.~\ref{fig:7} we compare the value of the heating force at the transonic point to the value given by Eq.~\eqref{eq:10} with $V=c_s$, as a function of the minimal distance $r_\mathrm{min}$. The analytic expectation is in reasonable agreement with the measured values. The ratio $\xi$ of the force value at the transonic point to its value at low Mach number is therefore:
\begin{equation}
  \label{eq:25}
  \xi\sim \frac{4}{\sqrt{\pi}\gamma}f\left(\frac{r_\textrm{min}V}{4\chi}\right).
\end{equation}
When the plume's size is much larger than the minimal length scale $r_\mathrm{min}$, this ratio is larger than one so that the force has a peak at the transonic point. Only when the minimal length scale is comparable to the plume's size does the peak vanish. In that case, by virtue of Eq.~\eqref{eq:23}, the asymptotic value at low Mach number is also significantly below its nominal value.

\begin{figure}
\includegraphics[width=.95\columnwidth]{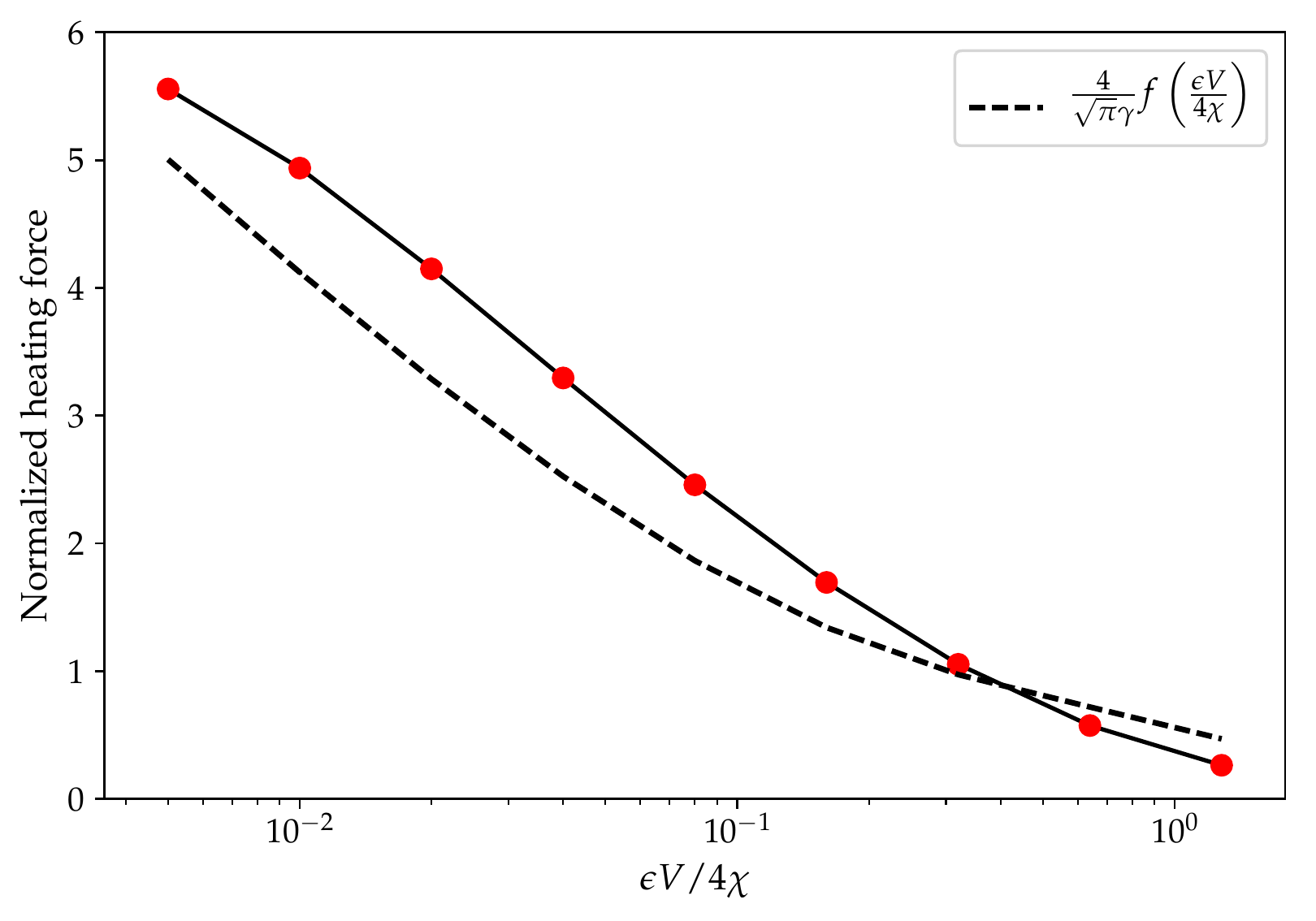}
 \caption{Heating force for $V=c_s$ at $t=50\tau$ normalized to its asymptotic value at low Mach number as a function of the ratio between softening length and resolution.}
 \label{fig:11}
\end{figure}

\subsection{Dynamical friction in presence of thermal diffusion}
\label{sec:dynam-frict-pres} In this section we study the force acting on a non-luminous perturber, when heat exchange can occur between neighbouring fluid elements, i.e. when we relax the adiabatic assumption of \citet{1999ApJ...513..252O}. In a first step we evaluate the correction to the force due to thermal diffusion using of linear perturbation theory and in a second step we check that expression using numerical simulations. In this whole section we therefore consider a cold perturber, the only difference with respect to the setup considered by \citet{1999ApJ...513..252O} being the introduction of thermal diffusion.

\subsubsection{Linear analysis of the cold thermal force in the limit of low Mach numbers}
\label{sec:linear-analysis-cold}
We start by linearising our governing equations [Eqs.~\eqref{eq:1} to~\eqref{eq:3}] which read:
\begin{eqnarray}
\label{eq:26}
\partial_t \rho'  +  \rho_0 \nabla \cdot \boldsymbol{v}' = 0 \\
\label{eq:27}
\partial_t\boldsymbol{v}' + \frac{\nabla{p'}}{\rho_0} = \nabla \Phi_p\\
\label{eq:28}
 \partial_t p'  +\gamma p_0 \nabla \cdot \boldsymbol{v}' + \frac{p_0}{\rho_0}\chi \Delta{\rho'} -\chi\Delta{p'}=0,
\end{eqnarray}
where prime quantities denote perturbations (e.g. $\rho'=\rho-\rho_0$). For the sake of legibility we hereafter drop the primes.
Using the same notation and same convention for Fourier transforms as  \citetalias{2017MNRAS.465.3175M} we get, upon Fourier transforming Eqs.~\eqref{eq:26} to~\eqref{eq:28} in space and time:
\begin{eqnarray}
\label{eq:29}
-i\omega\tilde{\rho} +  i \rho_0 k_j \tilde{v}_j = 0, j=1,2,3 \\
\label{eq:30}
-i\omega \tilde{v}_j + i k_j \frac{\tilde{p}}{\rho_0} = -i k_j \tilde\Phi_p\\
\label{eq:31}
-i\omega \tilde{p} +\gamma p_0 i k_j \tilde{v}_j - \frac{p_0}{\rho_0}\chi k^2 \tilde{\rho} +\chi k^2 \tilde{p}=  0.
\end{eqnarray} 
We can make use of Poisson's equation, which reads:
\begin{equation}
  \label{eq:32}
  \tilde\Phi_p=-\frac{4\pi G \tilde{\rho}_\mathrm{p}}{k^2},
\end{equation}
where $\rho_p(\boldsymbol{r},t)$ is the density of the perturber. Using Eqs.~\eqref{eq:29} and~\eqref{eq:30}, it follows that:
\begin{equation}
  \label{eq:33}
	\tilde{p}=\frac{4\pi G \rho_0\tilde\rho_p+\omega^2\tilde\rho}{k^2},
\end{equation}
which upon substitution in Eq.~\eqref{eq:31} yields the Fourier transform of the density perturbation:
\begin{equation}
\label{eq:34}
\tilde{\rho}=\frac{4\pi G\rho_0\left(\frac{i\omega}{k^2}-\chi\right)}{i\omega \left(c^2_s-\frac{\omega^2}{k^2}\right)+\chi \left(\omega^2-k^2c^2_s/\gamma\right)}\tilde\rho_p,
\end{equation}
where $c_s=\sqrt{\gamma p_0/\rho_0}$ is the adiabatic sound speed. When  $\chi=0$ this expression can be simplified into:
\begin{equation}
  \label{eq:35}
  	\tilde{\rho}_{\text{adi}}=\frac{4\pi G\rho_0}{k^2c^2_s-\omega^2}\tilde\rho_p,
\end{equation}
which is equivalent to Eq.~(4) of \citet{1999ApJ...513..252O}, and requires a time-dependent solution involving retarded Green's functions to give the drag force. In the general case, we can separate Eq.~\eqref{eq:34} into two components:
\begin{equation}
  \label{eq:36}
 	\tilde{\rho}=\tilde{\rho}_{\text{adi}}+\tilde{\rho}_{\text{thermal}},
\end{equation}
where      
\begin{equation}
  \label{eq:37}
 	\tilde{\rho}_{\text{thermal}}=\frac{4\pi G\rho_0 \chi k^2c^2_s(1-\gamma)/[\gamma(k^2c^2_s-\omega^2)]}{i\omega (c^2_s-\omega^2/k^2)+\chi(\omega^2-k^2c^2_s/\gamma)}\tilde\rho_p,  
\end{equation} 
is the new component that arises in the presence of thermal diffusion. This thermal component appears as the convolution product of the perturber's density $\rho_\mathrm{thermal}$ by a Green's function $K(\boldsymbol{r},t)$, the latter being the inverse Fourier transform of the ratio by which $\tilde\rho_p$ is multiplied in Eq.~\eqref{eq:37}. We now consider a perturber with a trajectory at constant speed $V$ along the $z$ axis that passes through $z=0$ at $t=0$. We evaluate the field $\rho_\mathrm{thermal}(\boldsymbol{r},t=0)$ in order to evaluate the force that it induces on the perturber in $\boldsymbol{r}=0$. The perturber's density field has the expression:
\begin{equation}
  \label{eq:38}
  \rho_p(x,y,z,t)=M\delta(x)\delta(y)\delta(z-Vt){\cal H}(-t),
\end{equation}
where ${\cal H}$ is Heaviside's step function, which we introduce to consider only the past ($t<0$) trajectory of the perturber. The convolution of $\rho_p$ by the Green's function $K$, which reads:
\begin{equation}
  \label{eq:39}
  \rho_\mathrm{thermal}(\boldsymbol{r},0)=\iiint d\boldsymbol{r'}\int dt'K(\boldsymbol{r}-\boldsymbol{r'},-t')\rho_p(\boldsymbol{r'},t')
\end{equation}
then takes the particularly simple form:
\begin{eqnarray}
  \rho_\mathrm{thermal}(x,y,z,0)&=&\int_{-\infty}^{+\infty}K(x,y,z-Vt',-t')M{\cal H}(-t')dt'\nonumber\\
  &=&\int_{-\infty}^0K\left(x,y,z-z',-\frac{z'}{V}\right)\frac MVdz'.
  \label{eq:40}
\end{eqnarray}
The force exerted on the perturber at $t=0$ is along the $z$ axis and has the expression:
\begin{eqnarray}
  F_\mathrm{thermal}^\mathrm{cold}&=&\!\!\!\!\!\!\iiint \frac{GM\rho_\mathrm{thermal}(x,y,z,0)z}{(x^2+y^2+z^2)^{3/2}}dx\,dy\,dz\nonumber\\
  &=&\!\!\!\!\!\!\int_{-\infty}^0\!\iiint \frac{GM^2K\left(x,y,z-z',-\frac{z'}{V}\right)z}{V(x^2+y^2+z^2)^{3/2}} dxdydz\,dz'
    \label{eq:41}
\end{eqnarray}
Only the norm $k$ of the wave vector features in Eq.~\eqref{eq:37} so that the Green's function has spherical symmetry.
The inner triple integral in Eq.~\eqref{eq:41} corresponds to the force exerted in $\boldsymbol{r}=0$ by a density perturbation equal to a Green's function centred on the location $(0,0,z')$. Since this function has spherical symmetry about that point, one may use Gauss' flux theorem to express this integral simply in terms of the mass enclosed within the sphere centred on $(0,0,z')$ with radius $-z'$. This mass can be easily worked out in the limit of a small Mach number. In that case the Green's function is evaluated at time $t=-z'/V$  for radii $|\boldsymbol{r}| < |z'|$, hence for radii $|\boldsymbol{r}|\ll tc_s$. In this limit we have in Fourier space $kc_s \gg \omega$ and we can simplify Eq.~\eqref{eq:37} as:
\begin{equation}
  \label{eq:42}
	\tilde{\rho}_\text{thermal}= \tilde K(k,\omega)\tilde\rho_p, 
      \end{equation}
      with
\begin{equation}
  \label{eq:43}
	\tilde K(k,\omega)= \frac{4\pi G\rho_0\chi (\gamma-1)/c^2_s}{\chi k^2 - i\gamma \omega }.
\end{equation} 
In real space we have therefore:
\begin{equation}
  \label{eq:44}
	K(\boldsymbol{r},t)= \frac{4\pi G\rho_0 \chi (\gamma-1)/(\gamma c^2_s)}{8(\pi\chi t/\gamma)^{3/2}}\exp^{\frac{-\gamma r^2}{4\chi t}}\mbox{~~for $r\ll tc_s$}.
\end{equation}
      In these conditions,
      the integral of $K(\boldsymbol{r},t)$ on a sphere of radius $R$ has the expression:
\begin{eqnarray}
	m(R,t)  &=& \int^{R}_0 4\pi r^2K(r,t)dr\nonumber\\
                                    &=& \frac{8\sqrt\pi G\rho_0\chi(\gamma-1)}{\gamma c^2_s}\left[-R\sqrt[]{\frac{\gamma}{4\chi t}}\exp^{\frac{-R^2\gamma}{4\chi t}}\right.\nonumber\\
  &&+ \left.\frac{\sqrt{\pi}}{2}\text{erf}\left(R\sqrt{\frac{\gamma}{4\chi t}}\right) \right].\label{eq:45}
\end{eqnarray}
Using Gauss's flux theorem we can therefore write:
\begin{equation}
  \label{eq:46}
  \iiint \frac{GMK\left(x,y,z-z',-\frac{z'}{V}\right)z}{(x^2+y^2+z^2)^{3/2}} dxdydz=-\frac{GMm(-z',-z'/V)}{z'^2}.
\end{equation}
Substituting this result in Eq.~\eqref{eq:41}, we eventually obtain:
\begin{eqnarray}
  F_\mathrm{thermal}^\mathrm{cold}&=&-\frac{2\pi^{\frac 12} G^2M^2\rho_0(\gamma-1)}{c_s^2}\!\!\int_0^{\infty}\left[-u^{\frac 12} e^{-u}+\frac{\sqrt\pi}{2}\text{erf}\left(u^{\frac 12}\right)\right]\frac{du}{u^2}\nonumber\\
  &=&-\frac{2\pi G^2M^2\rho_0(\gamma-1)}{c_s^2}
  \label{eq:47}
\end{eqnarray}
and, generalising to an arbitrary direction of motion:
\begin{equation}
  \label{eq:48}
F^\mathrm{cold}_\textrm{thermal}=-\frac{2\pi G^2M^2\rho_0(\gamma-1)}{c^2_s}\frac{\boldsymbol{V}}{V}
\end{equation}
This force is a drag force, like the dynamical friction in an adiabatic gas, as it is opposed to the motion. It has two remarkable properties: it is independent of the thermal diffusivity of the gas and of the perturber's velocity. The heating force has the same last property in the low Mach number regime.

It is also straightforward to realise, if we substitute in the analysis of section~3.1 of \citetalias{2017MNRAS.465.3175M} the convective derivative $V\partial_z$ with $\partial_t$, that the Fourier transform of the density perturbation arising from heat release on one hand, and the Fourier transform of the density perturbation arising from thermal diffusion [our Eqs.~\eqref{eq:42} and~\eqref{eq:43}] on the other hand, have same form, except for a change of sign. In the limit of low Mach numbers, the perturbation of density $\rho_\mathrm{thermal}$ has therefore the same shape as that of the hot, low-density plume associated to the heating force and corroborated by the numerical simulations presented in section~\ref{sec:low-mach-number}, except for a change of sign: it is here a cold and dense plume. This plume lagging the perturber, it is intuitive that it exerts an additional drag on it.

Following this analogy, we can further say that $\rho_\mathrm{thermal}$ has, in the low Mach number limit, the same form as the density perturbation that would be triggered by a massless heat sink with luminosity $-L_c$. The luminosity $L_c$ can be found by equating Eqs.~\eqref{eq:8} and~\eqref{eq:48}, which yields:
\begin{equation}
  \label{eq:49}
	L_c = \frac{4\pi G M \rho_0 \chi}{\gamma}.
\end{equation}
This luminosity has the same expression as that found in another context by \citet{2017MNRAS.472.4204M} for the cold thermal torques exerted on a low-mass planet on a circular orbit in a gaseous protoplanetary disc.

From the above we expect a perturber with a luminosity $L_c$ to be subjected to a heating force that nearly cancels the cold thermal force in the low Mach number regime. Such a perturber should therefore experience a drag force comparable to the force in an adiabatic gas.

\subsubsection{Numerical corroboration}
\label{sec:numer-corr}
\begin{figure*}
\includegraphics[width=\textwidth]{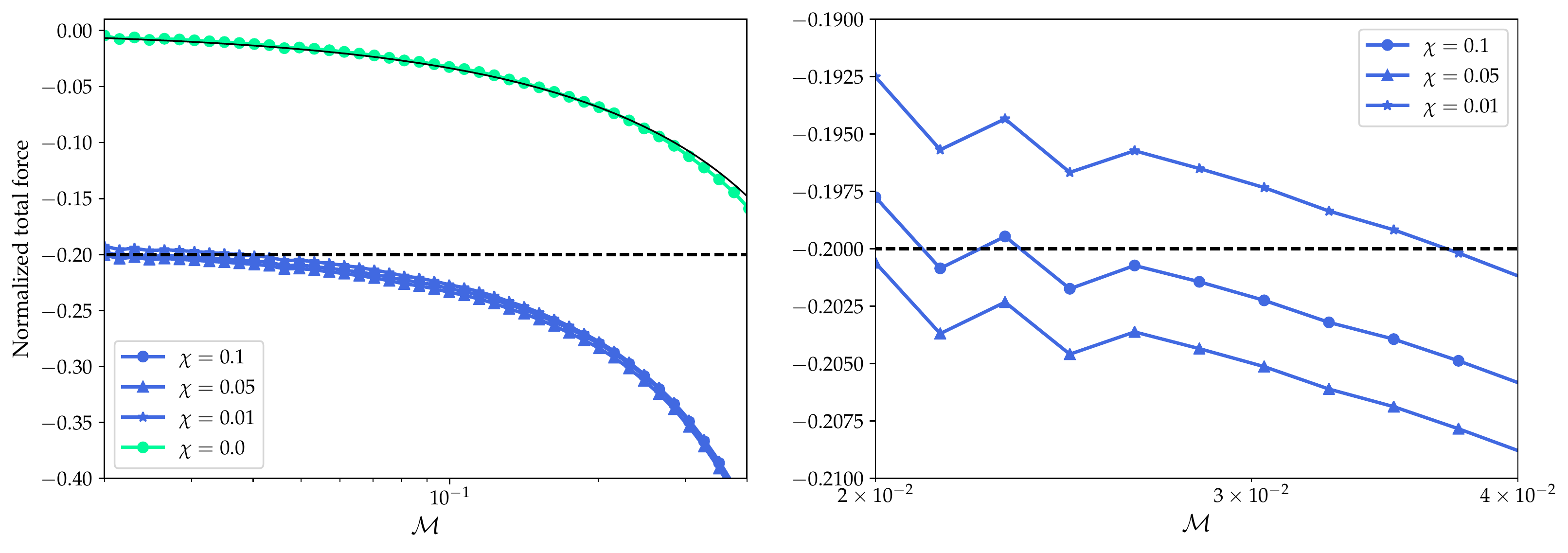}
 \caption{Net force on a cold perturber in an adiabatic gas (green curve) and in a gas with thermal diffusion (blue curves). The three blue curves are nearly superimposed on the left plot. The solid line represents the analytical expression of \citet{1999ApJ...513..252O} while the dashed line shows the asymptotic value of the net force in presence of thermal diffusion. The right plot shows a close-up for the diffusive cases at low Mach number regime. The forces are here normalized to $F_0=4\pi(GM)^2\rho_0/c_s^2$. The normalized force should therefore tend to $-(\gamma-1)/2=-0.2$ at vanishing velocity.}
 \label{fig:12}
\end{figure*}

In order to confirm the magnitude and direction of the cold thermal force, we ran simulations of a non-luminous perturber for different values of the thermal diffusivity ($\chi=0.01,0.05$ and $0.1$), spanning 44~values of the perturber's velocities in the subsonic regime ($\cal{M}$ $\in [0.02,0.7]$). The dimensions of the box as well as the final time were adjusted for $\chi=0.1$ (the most demanding case). Here we no longer restrict ourselves to the study of the heating force. It is therefore important to avoid the reflection of the acoustic sphere triggered at $t=0$ on the edges of the computational domain. We therefore chose $l=3200\lambda$ which prevents this reflection (except for the first two velocities where a bounce occurs near the end of the simulations). We have a ground level with size $64\times64\times128$, and $11$ levels of refinement with the first level covering a quarter of the ground level. Each level $\ell > 0$ has size $32\times32\times64$. With this system of meshes we expect to retrieve $\approx 99.1\%$, $98.4\%$ and  $92.3\%$ of the net force\footnote{Assuming the cold thermal force to have the same dependence on the largest and smallest scales as that given by Eq.~\eqref{eq:23} for the heating force.} respectively for the cases of $\chi=0.1,0.05$ and $0.01$. The results of these runs are shown in Fig.~\ref{fig:12}. They confirm the analytical expectations with a large degree of accuracy. The net force in runs with thermal diffusion is found to tend toward a finite, negative value in agreement to within a few percents with the prediction of Eq.~\eqref{eq:48}. This value is largely independent of the value of the thermal diffusivity. The minute variation from one run to the next with respect to this parameter can be attributed to changes in length scale coverage and to the onset of non-linearities in the flow near the perturber for the smallest value of $\chi$.

\begin{figure*}
  \centering
  \includegraphics[width=\textwidth]{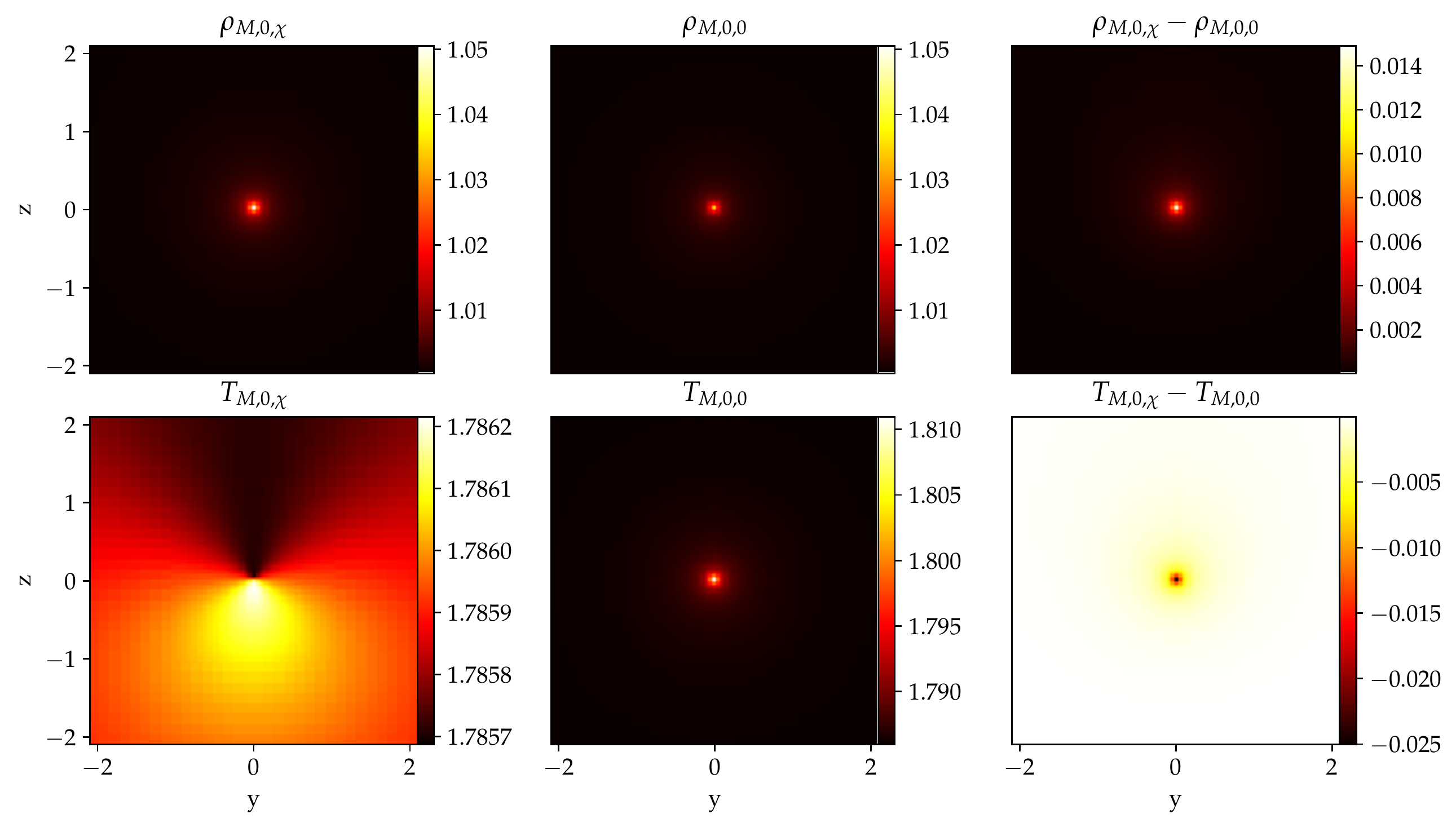}
  \caption{Density perturbation (top row) and temperature perturbation (bottom row) around a non-luminous perturber in a case with thermal diffusion (left column) and in the adiabatic case (middle column). It should be noted that the colour scale of the left bottom plot occupies a narrow interval, and that compared to the adiabatic case the temperature is nearly uniform around the perturber when thermal diffusion is present. The right column shows the difference between the diffusive and adiabatic cases.}
  \label{fig:13}
\end{figure*}

We also show the perturbations of density and temperature in the diffusive and adiabatic cases in Fig.~\ref{fig:13}, and the difference of these fields between these two cases. It confirms that the introduction of thermal diffusion induces a positive perturbation of density and a negative perturbation of temperature with respect to the adiabatic case.

\subsection{Net force in the general case}
\label{sec:net-force-general}
Having studied in separation the heating force and the cold thermal force, we are now in a position to study the net force on a perturber with mass $M$ and luminosity $L$ in a gas with thermal diffusion, when the flow is nearly linear. The net force is then given by the expression of Fig.~\ref{fig:1}.

\subsubsection{A simple expression of the thermal forces at low Mach number}
\label{sec:simple-expr-therm}
If we define: 
\begin{equation}
  \label{eq:50}
  F_0=\frac{4\pi(GM)^2\rho_0}{c_s^2}, 
\end{equation}
we can recast the expression of the thermal forces in the regime of low Mach number corresponding to Eqs.~\eqref{eq:48} and~\eqref{eq:8} respectively as: 
\begin{equation}
  \label{eq:51}
  F_\mathrm{thermal}^\mathrm{cold}=-\frac{\gamma-1}{2}F_0 
\end{equation}
and 
\begin{equation}
  \label{eq:52}
  F_\mathrm{heating}=\frac{\gamma-1}{2}\frac{L}{L_c}F_0, 
\end{equation}
so that 
\begin{equation}
  \label{eq:53}
F_\mathrm{thermal}=F_\mathrm{thermal}^\mathrm{cold}+F_\mathrm{heating}=\frac{\gamma-1}{2}\left(\frac{L}{L_c}-1\right)F_0. 
\end{equation}
We also note that the adiabatic force is, in terms of $F_0$, expressed as: 
\begin{equation}
  \label{eq:54}
  F_\mathrm{adi}=F_0{\cal M}^{-2}I 
\end{equation}
for all Mach numbers, where $I$ is given by Eqs.~(14) or~(15) of \citet{1999ApJ...513..252O}. 

\subsubsection{Numerical analysis}
\label{sec:numerical-analysis}
\begin{figure*}
\includegraphics[width=.95\textwidth]{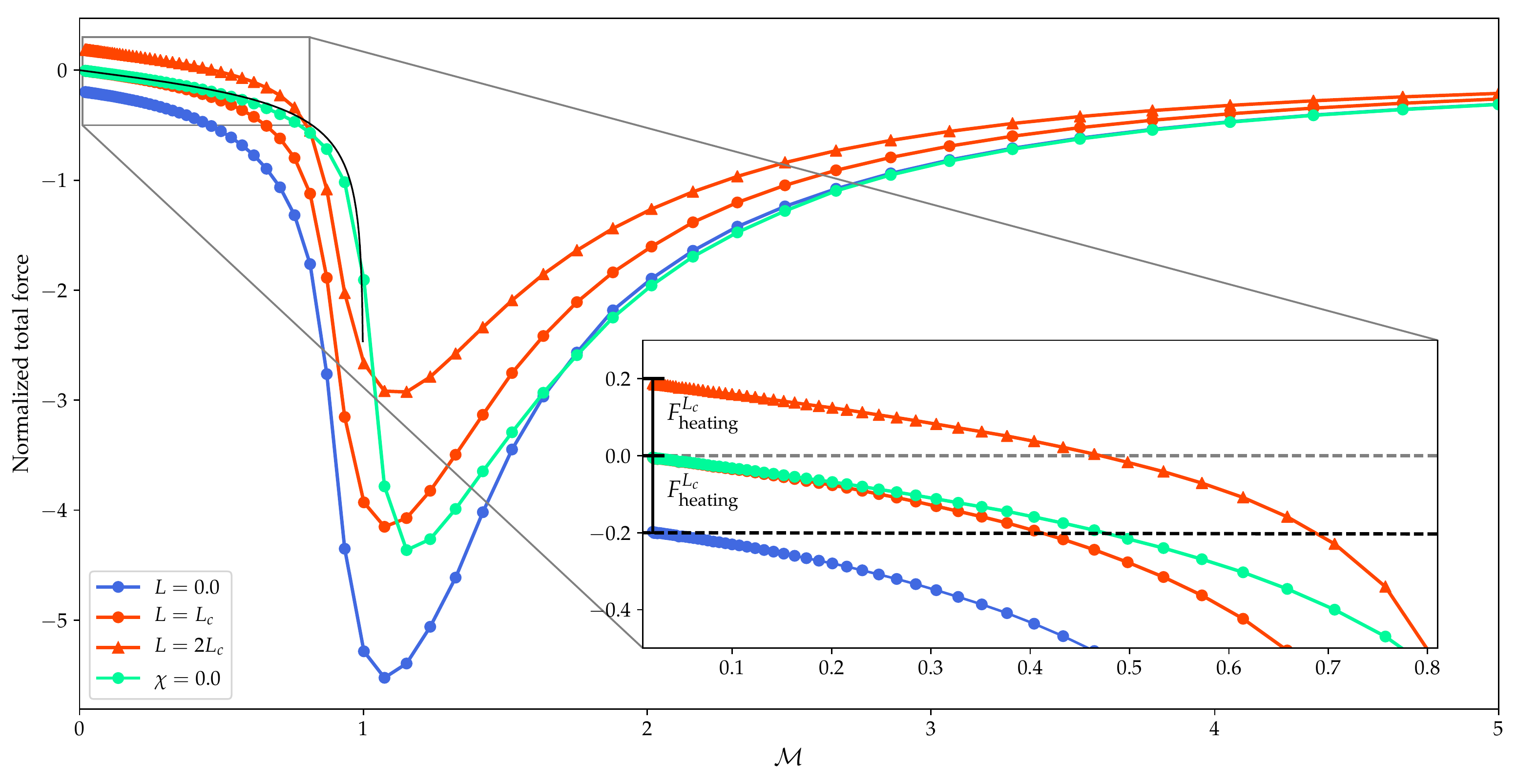}
\caption{Total force as a function of the Mach number in the adiabatic case ($\chi=0$) and  in a case with thermal diffusion for different luminosities. The inset plot shows that the adiabatic case and the case with thermal diffusion and $L=L_c$ coincide in the low Mach number limit. We also see that the runs with $L=2L_c$ (triangles) have a positive net force at low Mach number.}
 \label{fig:14}
\end{figure*}

\begin{figure*}
\includegraphics[width=.95\textwidth]{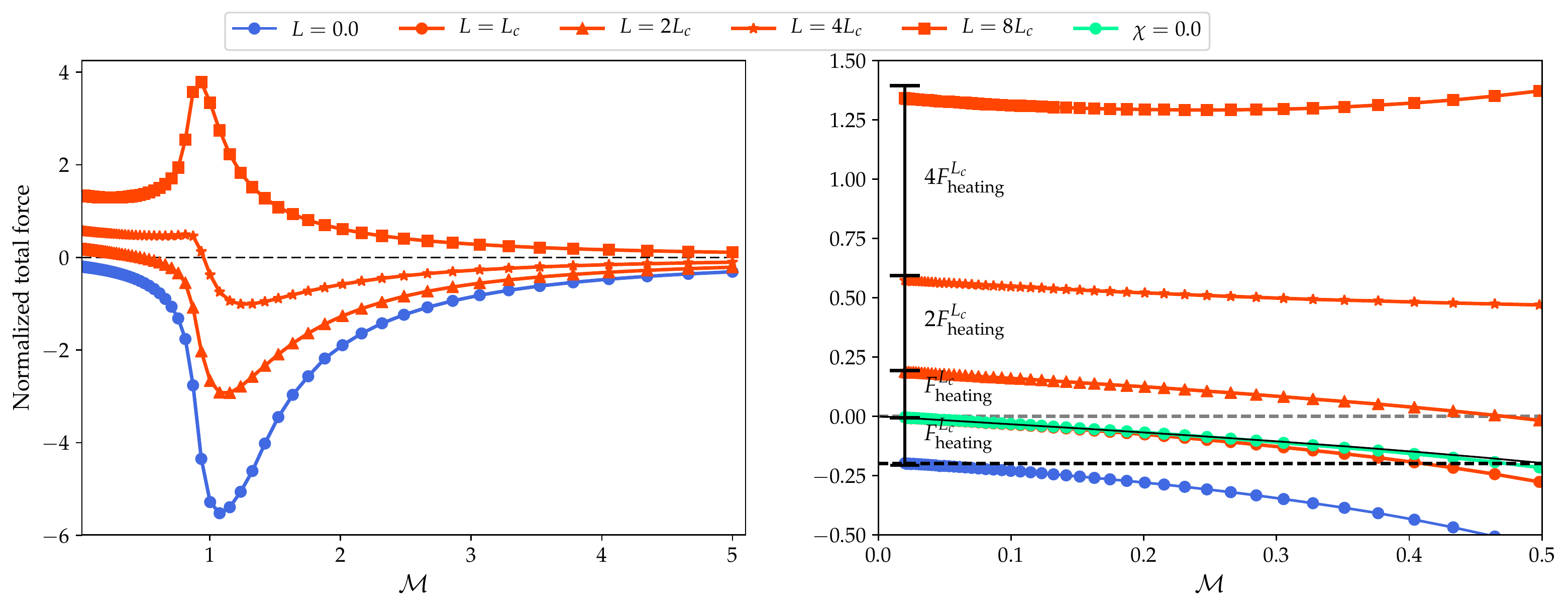}
 \caption{Net force as a function of the Mach number for various luminosities. The right plot presents a close up of the results in the subsonic regime and shows that the heating force scales with the luminosity up to the largest value ($8L_c$) used in the calculations.}
 \label{fig:15}
\end{figure*}

\begin{figure}
\includegraphics[width=.95\columnwidth]{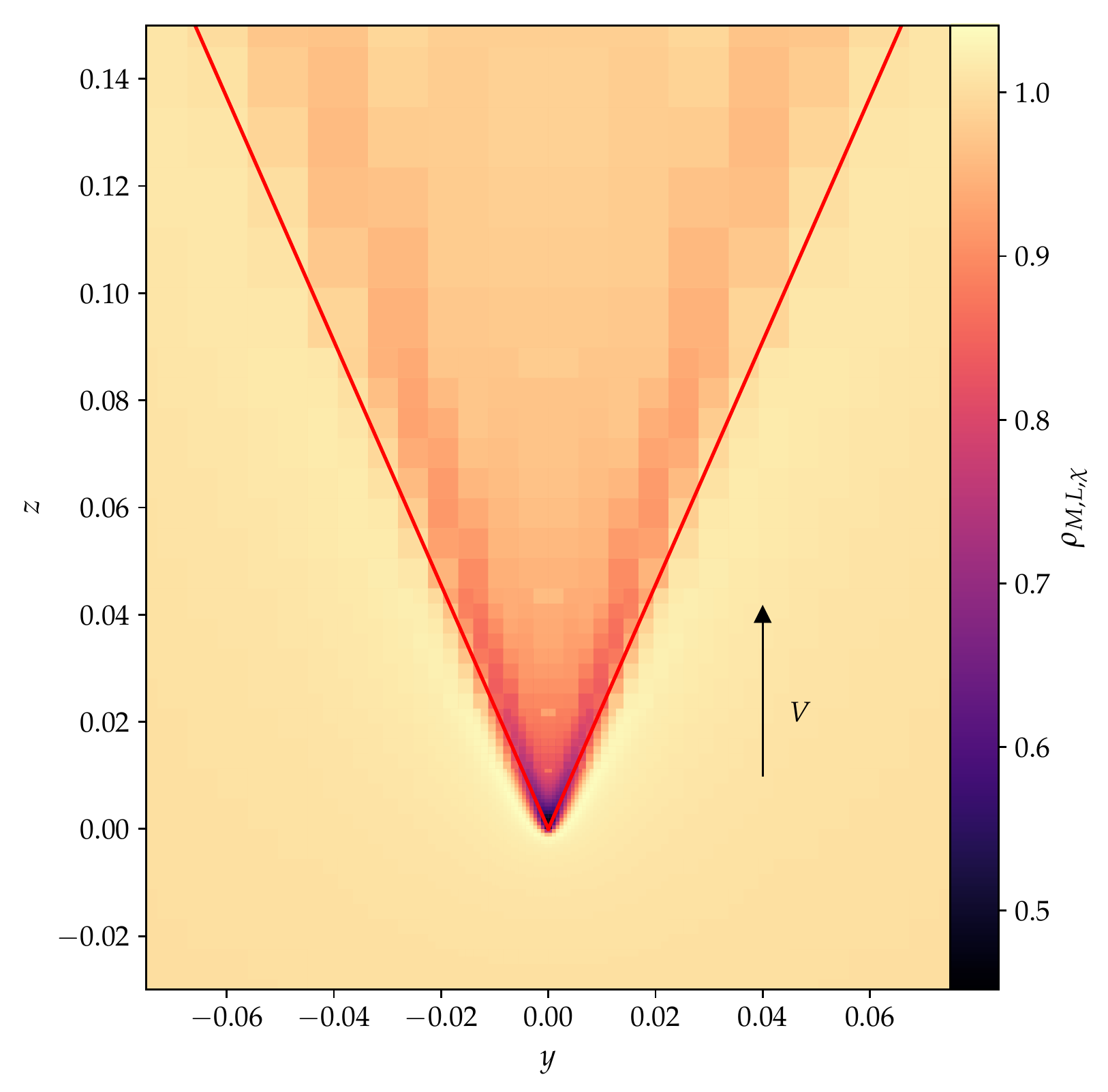}
 \caption{Snapshot of the density for ${\cal M}=2.48$ and $L=8L_c$. We see that the interior of the Mach cone is less dense than its surroundings, consistent with the net thrust observed in Fig. \ref{fig:15}.}
 \label{fig:16}
\end{figure}

In Fig.~\ref{fig:14} we present the results for different sets of simulations using  11 levels of refinement to reach $r'_{min}=0.01$ in a box of $r'_{max}=1600$, therefore retrieving $99\%$ of the net heating force. We show the total force for the adiabatic case ($\chi=0$), and for cases with thermal diffusivity $\chi=0.1$ for three different values of the luminosity: $L=0$, $L=L_c$ and $L=2L_c$. We confirm our expectation that the net force nearly coincides with the adiabatic value of \citet{1999ApJ...513..252O} when $L=L_c$ in the low Mach number regime (see last paragraph of section~\ref{sec:linear-analysis-cold}).
Our analysis of the cold thermal force of section~\ref{sec:dynam-frict-pres} was limited to the low Mach number regime. We have not worked out an expression of that force in the supersonic regime. However, our simulations seem to indicate that in this regime the cold thermal force has a negligible role, since the net force without heating (blue curve in the electronic version of this manuscript) tends to the adiabatic force (green curve in the electronic version of this manuscript).  We also find that the net force is positive, and corresponds to a thrust on the perturber, when the luminosity is in excess of $L_c$ and the Mach number is sufficiently low (e.g. below $\approx 0.48$ when $L=2L_c$).
\citet{2017MNRAS.465.3175M} argued that, since both the adiabatic force and the heating force follow a $V^{-2}$ law in the supersonic regime and have opposite signs, the net force may be positive at all speeds in the supersonic regime when the perturber's luminosity is sufficiently large. Their analysis disregarded the cold thermal force and assumed the force on a non-luminous perturber to coincide with the adiabatic estimate of \citet{1999ApJ...513..252O}. Our present findings, however, suggest that the cold thermal force is negligible in the supersonic regime, so that their expectation should be correct. We have therefore undertaken additional sets of calculations with higher luminosities, presented in Fig.~\ref{fig:15}. We see that for a luminosity $L=4L_c$, the net force is positive essentially over the whole subsonic domain, while it is always positive for $L=8L_c$, for the set of parameters adopted. This behaviour is consistent with the density profile depicted by Fig. \ref{fig:16} in which we observe a deficit of material behind the perturber inside the Mach cone. 

\section{Discussion and conclusions}
\label{sec:disc-concl}
We have conducted extensive numerical simulations in order to check the impact of thermal effects on the dynamical friction onto a massive perturber, both with and without intrinsic luminosity (i.e. with or without radiative feedback). Owing to the wide interval of length scales required to capture accurately thermal effects, we had to resort to nested meshes, that we implemented in the GPU version of the FARGO3D code. As a side feature, the present manuscript is intended to stand as a concise reference to our nested mesh implementation.

Our simulations confirmed with a high degree of accuracy the analytic predictions of \citet{2017MNRAS.465.3175M}. Namely, when the perturber is luminous, the extra force arising from the density perturbation due to the radiative feedback, or heating force, is found to behave as expected in the regimes of low and high Mach numbers. In the regime of low Mach numbers, it tends toward a finite value when the velocity tends to zero. This somehow counter-intuitive behaviour occurs owing to the following scalings. The region that gives rise to the net force has a typical size $\sim\lambda$ and therefore a volume $V \propto \lambda^3$. The perturbation of density in this region scales as $\rho \propto {\lambda}^{-1}$, therefore leading to a perturbation of mass $\delta M \propto \lambda^2$. It then follows that the net force arising from this volume scales as $GM \delta M/\lambda^2$ and  is therefore independent of the cut-off distance $\lambda$. The only effect of changing the velocity on the perturbation of density is to change the scale $\lambda$. It therefore follows that the heating force does not depend on the velocity. However, as the perturber's velocity tends to zero, the cut-off distance increases (as diffusion can go a long way against the wind) and it takes more time to establish the perturbed density field. In the low Mach number regime, the acceleration due to the heat release is a function of $t/\tau$ only, where $\tau$ is defined by Eq.~\eqref{eq:9}. This function tends toward the asymptotic value of Eq.~\eqref{eq:8} and $\tau$ tends to infinity when the Mach number vanishes.

In the high Mach number regime the heating force decays as $V^{-2}$, and exhibits a logarithmic dependence on the smallest length scale. In all cases the heating force is along the direction of motion and yields a thrust on the perturber.

The present work also shows that the gravitational drag on a non-luminous perturber, when thermal diffusion is present, is significantly at odds with the expression worked out by \citet{1999ApJ...513..252O} for an adiabatic gas. The offset between the two expressions, that we dub the cold thermal force, shares a number of similarities with the heating force in the subsonic regime. In particular, it tends towards a finite value in the limit of a vanishing velocity. The interpretation of this new force component is very similar to that of the cold thermal torque discussed by \citet{2017MNRAS.472.4204M} in a different context (that of low mass protoplanets in a gaseous protoplanetary disc). The introduction of thermal diffusion flattens out the temperature peak that would otherwise surround the perturber if the gas were adiabatic. Allowing thermal diffusion is therefore tantamount to introducing an additional, negative perturbation of temperature in the vicinity of the perturber. Pressure balance in the low Mach number regime then implies that this negative perturbation is accompanied by a positive perturbation of density. \citet{2017MNRAS.472.4204M} argues that when the perturber has a very subsonic velocity with respect to the ambient gas,  the introduction of thermal diffusion yields a perturbation on the gas that is equivalent to the perturbation that would be induced by a massless heat sink with luminosity $-L_c$, where $L_c$ is given by Eq.~\eqref{eq:49}, regardless of the geometry of the background flow (whether it is sheared, like in a protoplanetary disc, or not, like in the situation considered here). This is exactly what we find in the present analysis and the simulations presented here constitute the first quantitative confirmation of this effect. One potentially important consequence of this effect is the eccentricity damping of cold embryos in protoplanetary discs. It should be much more vigorous than expected from adiabatic calculations. This has been noticed in numerical simulations by \citet{2017arXiv170401931E}, but not in a quantitative manner, owing to resolution constraints.

The cold thermal force being finite at vanishing velocity, it behaves like solid or dry friction, and halts the perturber in a finite time. As the response time of the plume tends to infinity in the limit of vanishing velocity, this can only be true, however, if the stopping distance is a fraction of the cut-off distance $\lambda$ for the initial velocity. It is easy to check that a sufficient condition for that is $L_c \gg c_s^5/G$.

The finite, negative force in the limit of a vanishing velocity for a non-luminous perturber that we found here had not been considered by \citet{2017MNRAS.465.3175M}. Their conclusions need therefore to be amended. A perturber in slow motion is accelerated only if its luminosity is larger than $L_c$. Otherwise, the net force acting on it is negative, and its velocity decays. For the two fiducial disc models considered by these authors, the critical luminosity for a planetary embryo with mass $\chi c_s/G$ coincides with the accretion luminosity for a mass doubling time of roughly $100$~kyrs. For the disc parameters considered by \citet{2015Natur.520...63B} and \citet{2017arXiv170401931E}, the accretion luminosity for a mass doubling time of $100$~kyrs is twice the critical luminosity of an embryo with critical mass $\chi c_s/G$. Low mass embryos subjected to fast pebble accretion \citep{2012A&A...544A..32L}, for which mass doubling times largely shorter than $100$~kyrs can be expected, should therefore be subjected to a heating force largely in excess of the cold thermal force.

Although we have not worked out an analytical expression of the cold thermal force in the supersonic regime, our numerical experiments suggest it to be negligible. A tentative explanation could be as follows: the cold force primarily arises from the flattening, by thermal diffusion, of the temperature peak associated to a nearly hydrostatic envelope in the slow regime. No such envelope can build up in the supersonic regime, so that the introduction of thermal diffusion in that regime induces much less significant perturbations in the perturber's vicinity.

We have found that a sufficiently large luminosity of the perturber warrants a net positive force at all Mach numbers.  We comment that the gravitationally induced drag diverges logarithmically as time goes on in the supersonic regime while the heating force remains bound, so that ultimately the net force will revert sign in this regime and will become negative, unless the medium is bound, such as in a thin protoplanetary disc \citep{2011ApJ...737...37M}. In that case, the net force may remain positive at all times and the perturber's velocity increases as $t^{1/3}$. The hot plume's size being inversely proportional to the perturber's velocity, the acceleration ceases when the plume size becomes comparable to the perturber's physical size, or, if thermal diffusion is effected by radiative transfer, when it becomes comparable to the photons' mean free path, whichever occurs first.

In the present analysis we have restricted ourselves to weakly perturbed flows, as we have not tried to resolve the flow within the Bondi sphere of the perturber. High resolution simulations resolving the flow in the Bondi sphere and evaluating its impact on the thermal forces will be presented in a forthcoming work.




\bibliographystyle{mnras}
\bibliography{biblio} 



\appendix

\section{Sub-cycling criterion}
\label{sec:sub-cycl-crit}
In order to show how the time stepping schedule can be optimised, we take as an example a low-mass planet embedded in a nearly Keplerian protoplanetary disc, so that the gas is nearly at rest in the planet's vicinity. The standard sub-cycling technique may not be the most appropriate in this situation: owing to the shear, the flow velocity (which imposes the strongest constraint on the time step, in a geometrically thin disk) is smaller for the higher resolution meshes, which have a smaller radial extent from corotation. For the sake of definiteness consider a mesh with one level of refinement only in the planet vicinity. Assume that the CFL condition, evaluated separately on the coarse mesh, yields a time step
limit $\delta t_0$, and evaluated separately on the fine mesh, yields
a time step limit $\delta t_1=K\delta t_0$, where we assume
$1/2<K<1$. For simplicity we assume that the fine mesh has
the same number of zones as the coarse one, so that an integration on
either mesh has same computational cost $C_0$. We can now consider two possibilities:
\begin{itemize}
\item We perform two timesteps on the finer mesh for one timestep on
  the coarser mesh (sub-cycling). The global time step is therefore
  limited by $\delta t_0$, since $K>1/2$, and the total cost of an
  integration on the whole domain is $3C_0$.
\item We use the same time step for both meshes (flat case). The global time step is therefore limited by $\delta t_1$ and the total cost of an integration on the whole domain is $2C_0$.
\end{itemize}
We naturally wish to maximise the time elapsed in the simulation for a
given amount of wall clock time, so we search the procedure that has
the largest time step to cost ratio. For the first case contemplated
here, it is $\delta t_0/3C_0$, whereas for the second case it is
$K\delta t_0/2C_0$. We see that if $K>2/3$, the flat procedure
(no sub-cycling) is more efficient than the other one. As
anticipated above, it is in the case where the timestep limit does not
differ much on the fine and coarse meshes that a flat procedure is
desirable. This simple example can be generalised to a case of $n$
levels of refinement, each corresponding to a computational cost
$(C_{\ell})_{0\leq \ell<n}$, and each having an individual time step limit $(\delta t_{\ell})_{0\leq \ell <n}$. A time step at a given level $\ell$ can correspond either to one (flat case) or two (sub-cycling) time steps at level $\ell+1$, independently of how many timesteps are performed at level $\ell+2$ for one time step at level $\ell+1$. We therefore construct a sequence $(\xi_\ell)_{0\leq \ell < n-1}$, such that $\xi_\ell=1$ if the same time step is used for levels $\ell$ and $\ell+1$, and $\xi_\ell=2$ otherwise.
The timestep on the whole domain that satisfies the CFL condition
for all levels is:
\begin{equation}
  \label{eq:55}
  \Delta t = \min\left[\delta t_\ell \prod_{m=0}^{m<\ell}\xi_m\right]_{0\leq \ell<n},
\end{equation}
where the product is meant to be $1$ when $\ell=0$, and the cost of one
full time step is
\begin{equation}
  \label{eq:56}
  C=\sum_{\ell=0}^{n-1}\left(C_\ell\prod_{m=0}^{m<\ell}\xi_m\right).
\end{equation}
We seek the sequence $(\xi_m)_{0\leq m<n-1} \in \{1,2\}^{n-1}$ for which the ratio $\Delta t/C$ is maximal. A given sequence can be regarded as a binary word of length $n-1$ composed of 1's and 2's. We therefore have to evaluate Eqs.~(\ref{eq:55}) and~(\ref{eq:56}) for the $2^{n-1}$ possible cases. This optimisation is performed at the beginning of each coarse time step, with a negligible computational overhead. For the setup considered in the present paper, this technique falls back to using $\xi_\ell=2$ for  $0 \leq \ell < n$.


\bsp	
\label{lastpage}
\end{document}